\documentclass[superscriptaddress,twocolumn,showpacs,amsmath,amssymb,titlepage]{revtex4}
\usepackage{graphicx}
\usepackage{dcolumn}
\usepackage{bm}
\usepackage{latexsym}
\begin{document} 
\title{Machines of life: catalogue, stochastic process modeling, \\
probabilistic reverse engineering and the PIs- from Aristotle to Alberts} 
\author{Debashish Chowdhury}
\email{debch@iitk.ac.in}
\affiliation{Department of Physics, Indian Institute of Technology, 
        Kanpur 208016, India} 
\affiliation{Mathematical Biosciences Institute, The Ohio State University,
Columbus, OH 43210, USA.}
\date{\today}
\begin{abstract}
Molecular machines consist of either a single protein or a macromolecular 
complex composed of protein and RNA molecules. Just like their macroscopic 
counterparts, each of these nano-machines has an engine that ``transduces'' 
input energy into an output form which is then utilized by its coupling to 
a transmission system for appropriate operations. The theory of heat engines, 
pioneered by Carnot, rests on the second law of equilibrium thermodynamics. 
However, the engines of molecular machines, operate under isothermal 
conditions far from thermodynamic equilibrium. Moreover, one of the possible 
mechanisms of energy transduction, popularized by Feynman and called Brownian 
ratchet, does not even have any macroscopic counterpart. But, {\it molecular 
machine  is not synonymous with Brownian ratchet}; a large number of molecular 
machines actually execute a noisy power stroke, rather than operating as 
Brownian ratchet. The man-machine analogy, a topic of intense philosophical 
debate in which many leading philosophers like Aristotle and Descartes 
participated, was extended to similar analogies at the cellular and 
subcellular levels after the invention of optical microscope. The idea of 
molecular machine, pioneered by Marcelo Malpighi, has been pursued 
vigorously in the last fifty  years. It has become a well established topic 
of current interdisciplinary research as evident from the publication of a 
very influential paper by Alberts towards the end of the twentieth century. 
Here we give a non-technical overview of the strategies for (a) stochastic 
modeling of mechano-chemical kinetic processes, and (b) model selection 
based on statistical inference drawn from analysis of experimental data.
It is written for non-experts and from a broad perspective, showing 
overlapping concepts from several different branches of physics and from 
other areas of science and technology.

\end{abstract}

\maketitle


\newpage 
\begin{widetext}
\begin{verbatim} 
Table of contents: 

1. Introduction 
II. A catalogue: typical examples of molecular machines and fuels 
    II.A. Fuels for molecular machines 
    II.B. Cytoskeletal motors 
          II.B.1. Porters 
          II.B.2. Machines for chipping filamentous tracks 
          II.B.3. Contractility: motor-filament crossbridge and 
                  collective dynamics of sliders and rowers  
          II.B.4. Push and pull of cytoskeletal filaments: nano-pistons and nano-springs
    II.C. Machines for synthesis, manipulation and degradation of macromolecules of life 
          II.C.1. Membrane-associated machines for macromolecule 
                  translocation: exporters, importers and packers 
          II.C.2. Machines for degrading macromolecules of life 
          II.C.3. Machines for template-dictated polymerization 
          II.C.4. Unwrappers and unzippers of packaged DNA: chromatin remodellers and helicases 
    II.D. Rotary motors 
    II.E. Processivity, duty ratio, stall force and dwell time 
    II.F. Fundamental general questions 
III. Motoring in a viscous fluid: from Newton to Langevin 
     III.A. Newton's equation: deterministic dynamics 
     III.B. Langevin equation: stochastic dynamics 
            III.B.1. Motoring in a ``sticky'' medium: Purcell's idea 
            III.B.2. Motoring under random impacts from surroundings: Brownian force 
IV. Energy transduction by molecular machines: from Carnot to Feynman 
    IV.A. Molecular machines are run by isothermal engines 
    IV.B. Inadequacy of equilibrium thermodynamics 
    IV.C. Inadequacy of endo-reversible thermodynamics 
    IV.D. Domain of non-equilibrium statistical mechanics 
    IV.E. Defining efficiency: from Carnot to Stokes 
    IV.F. Noisy Power stroke and Brownian ratchet 
    IV.G. Physical realizations of Brownian ratchet in molecular motors 
V. Mathematical description of mechano-chemical kinetics: 
   continuous landscape versus discrete networks 
   V.A. Motor kinetics as wandering in a time-dependent mechano-chemical free-energy landscape 
   V.B. Motor kinetics as wandering in the time-dependent real-space potential landscape 
   V.C. Motor kinetics as a jump process in a fully discrete mechano-chemical network 
   V.D. Balance conditions for mechano-chemical kinetics: cycles and flux 
VI. Solving forward problem by stochastic process modeling: from model to data 
    VI.A. Average speed and load-velocity relation 
    VI.B. Jamming on a crowded track: flux-density relation 
    VI.C. Information processing machines: fidelity versus power and efficiency 
    VI.D. Beyond average: dwell time distribution 
          VI.D.1. DTD for a motor that never steps backward 
          VI.D.2. Conditional DTD for motors with both forward and backward stepping 
VII. A summary of experimental techniques: ensemble versus single-machine
VIII. Solving inverse problem by probabilistic reverse engineering: from data to model 
     VIII.A. Frequentist versus Bayesian approach 
            VIII.A.1. Maximum-likelihood estimate 
            VIII.A.2. Bayesian estimate 
            VIII.A.3. Hidden Markov models 
     VIII.B. Extracting FP-based models from data? 
IX. Overlapping research areas 
      IX.A. Symmetry braeking: directed motility and cell polarity 
      IX.B. Self-organization and pattern formation: assembling 
              machines and cellular morphogenesis 
      IX.C. Dissipationless computation: polymerases as ``tape-copying Turing machines'' 
      IX.D. Enzymatic processes: conformational fluctuations, static and dynamic disorder 
      IX.E. Applications in biomimetics and nano-technology 
X. Concept of biological machines: from Aristotle to Alberts 
XI.  Summary and outlook 
\end{verbatim} 
\end{widetext}

\newpage

\section{Introduction}

Cell is the structural and functional units of life. How ``active'' is the interior of a living cell? 
Imagine an under water ``metro city'' which is, however, only about 
$10 \mu$m long in each direction! In this city, there are ``highways'' 
and ``railroad'' tracks on which motorized ``vehicles'' transport 
cargo to various destinations. It has an elaborate mechanism of 
preserving the integrity of the chemically encoded blueprint of the 
construction and maintenance of the city. The ``factories'' not only 
supply their products for the construction and repair works, but 
also manufacture the components of the machines. This eco-friendly 
city re-charges spent ``chemical fuel'' in uniquely designed ``power 
plants''. This city also uses a few ``alternative energy'' sources 
in some operations. Finally, it has special ``waste-disposal plants'' 
which degrade waste into products that are recycled as raw materials 
for fresh synthesis. All the automated processes in this high-tech 
micro-city are run by {\it nano-machines} \cite{frank11}. This is not 
the plot of a science fiction, but a dramatized picture of the dynamic 
interior of a cell. 

A molecular machine is either a single protein or a macromolecular 
complex consisting of proteins and RNA molecules \cite{chowdhury09}. 
Just like their macroscopic counterparts, these nano-machines take 
an input (most often, chemical energy) and it transduces input 
energy into an output. If the output is mechanical work the machine 
is usually referred to as a molecular motor \cite{howard01a,schliwa03,fisher07}. 
Similarly, a cell can also be regarded as a micron-size ``energy 
transforming device''. The concept of ``machine'' is not restricted 
only to subcellular or cellular levels of biological organization. 
In ancient times, philosophers proposed the concept of ``living 
machine'' to describe a whole animal, including a man; we'll discuss 
the evolution of this concept towards the end of this article. 

Not only animals, but even plants also move in response to external 
stimuli \cite{skotheim05,martone10} and movements take place in 
plants at all levels- from whole plant and plant cell to the 
subcellular level. 
I cannot resist the temptation of quoting Thomas Huxley's poetic 
description of cytoplasmic streaming \cite{shimmen07}, in particular, and of 
intracellular motor traffic, in general, in plants \cite{huxley1869}: 
``{\it ..the wonderful noonday silence of a tropical forest is, after all, due 
only to the dulness of our hearing; and could our ears catch the murmur of 
these tiny maelstroms, as they whirl in the innumerable myriads of living 
cells which constitute each tree, we should be stunned, as with the roar of 
a great city}''. 

In this article, however, we focus attention almost exclusively on 
molecular machines that drive subcellular processes within living 
cells. The processes driven by molecular machines include not only 
intracellular motor transport, but also manipulation, polymerization 
and degradation of the bio-molecules \cite{cozzarelli06a,stark10}. 
Explaining the physical principles that govern these machine-driven 
processes will bring us closer to the ultimate goal of biological 
physics: understanding ``what is life'' in terms of the laws of 
physics (and chemistry).

Biomolecular machines operate in a domain where the appropriate units of 
length, time, force and energy are, {\it nano-meter}, {\it milli-second}, 
{\it pico-Newton} and $k_BT$, respectively ($k_B$ being the Boltzmann 
constant and $T$ is the absolute temperature). Aren't the operational 
mechanism of molecular machines similar to their macroscopic counterparts 
except, perhaps, the difference of scale? NO. In spite of the striking 
similarities, it is the differences between molecular machines and their 
macroscopic counterparts that makes the studies of these systems so 
interesting from the perspective of physicists. 

This article is a brief guide for a beginner embarking on an exploration 
of the exciting frontier of research on molecular machines. It begins 
with a catalog of the known molecular machines and a list of fundamental 
questions on their operational mechanism that the explorer should try to 
address. Like any travel guide, this article also cautions the explorer 
about the counter-intuitive phenomena that await him/her in this new 
territory of his/her exploration where he/she may need new sophisticated 
technical tools that were not needed for handling macroscopic machines. 
For the explorer, this article also charts a tentative road map along with 
a summary of the mathematical strategies that he/she might use to make 
progress in this frontier territory. Excursions to some of the listed 
border areas that overlap with areas of research in other branches of 
science and engineering could be enjoyable and intellectually rewarding 
for the explorer. For any explorer, it is good to know the experience 
of the pioneers. The final section, on the evolution of the concept of 
biological machines, is likely to be enjoyed as a dessert by physicists, 
and as a fresh food for thought by historians and philosophers of science.

\section{A catalogue: typical examples of molecular machines and fuels}

Based on the mode of operation, the biomolecular machines can be 
divided broadly into two groups. {\it Cyclic machines} operate in 
repetitive cycles in a manner that is very similar to that of the 
cyclic engines which run our cars. In contrast, some other molecular 
machines are {\it one-shot} machines that exhaust an internal source 
of free energy in a single round. Force exerted by a compressed spring, 
upon its release, is a typical example of a one-shot machine.

\subsection{Fuels for molecular machines}

The most common way of supplying energy to a natural nanomachine is
to utilize the chemical energy (or, more appropriately, free energy)
released by a chemical reaction. Most of the machines use the so-called
high-energy compounds- particularly, nucleoside triphosphates (NTPs)-
as an energy source to generate the mechanical energy required
for their directed movement. The most common chemical reaction is the
{\it hydrolysis} of ATP (ADP): $ ATP \rightarrow ADP + P_i$.
Some other high-energy compounds can also supply input energy; one 
typical example being the hydrolysis of Guanosine Triphosphate (GTP)
to Guanosine Diphosphate (GDP).
Under normal physiological conditions, hydrolysis of ATP is extremely 
slow. Most of the molecular motors fuelled by ATP function also as 
ATPase enzyme (i.e., catalyzes hydrolysis of ATP) thereby speeding up 
the energy-supplying reaction by several orders of magnitude.

If a cyclic machine runs on a specific chemical fuel then the spent
fuel must be removed as waste products and fresh fuel must be supplied
to the machine. Fortunately, normal cells have machineries for 
recycling waste products to manufacture fresh fuel, e.g., synthesizing 
ATP from ADP. This raises an important question: since ATP is a 
higher-energy compound than ADP, how are the ATP-synthesizing machines 
driven to perform this energetically ``uphill'' task? Fortunately, 
chemical fuel is not the only means by which input energy can be 
supplied to intracellular molecular machines.

A cell gets its energy from external sources. It has special machines 
to convert the input energy into some ``energy currency''. For example, 
chemical energy supplied by the food we consume is converted into an 
electro-chemical potential $\Delta \mu$ that not only can be used to 
synthesize ATP, but can also directly run some other machines. 
In plants similar proton-motive forces are 
generated by machines which are driven by the input sunlight.  
Only hydrogen ion ($H^+$), i.e., a proton, and sodium ion ($Na^+$) 
are used in the kingdom of life to create the electro-chemical 
potentials, i.e., it used either a proton-motive force (PMF) or a 
sodium-motive force (SMF) as an energy currency.    

The advantages of using light, instead of chemical reaction, as the 
input energy for a molecular motor are as follows: (i) light can be 
switched on and off easily and rapidly, (ii) usually, no waste 
product, which would require disposal or recycling, is generated.

Thus, study of molecular machines deals with two complementary 
aspects of bioenergetics: (a) conversion of energy input from the 
external sources into the energy currency of the cell, and (b) 
conversion of the energy currency to drive various active other 
active processes.

\subsection{Cytoskeletal motors}

The cytoskeleton of a cell is the analogue of the human skeleton 
\cite{howard01a}. However, it not only provides mechanical strength 
to the cell, but its filamentous proteins also form the networks of 
``highways'' (or, ``tracks'') on which cytoskeletal motor proteins 
\cite{howard01a,schliwa03} can move.  Filamentous actin (F-actin) 
and microtubules (MT) which serve as tracks are ``polar'' in the 
sense that the structure and kinetics of the two ends of each 
filament are dissimilar.

\begin{table}
\begin{tabular}{|c|c|c|} \hline
Motor superfamily & Filamentous track & Minimum step size \\\hline
Myosin & F-actin & 36 nm \\ \hline
Kinesin & Microtubule & 8 nm \\ \hline
Dynein & Microtubule & 8 nm \\ \hline
\end{tabular}
\caption{Superfamilies of motor proteins and the corresponding tracks.
 }
\label{table-cytocomp}
\end{table}

The superfamilies of cytoskeletal motors and the corresponding filamentous 
tracks are listed in table \ref{table-cytocomp}. 
Every superfamily can be further divided into families. Members of every 
family move always in a particular direction on its track; for example, 
kinesin-1 and cytoplasmic dynein move towards + and - end of MT, 
respectively. Similarly, myosin-V and myosin-VI move towards the + and 
- ends of  F-actin, respectively. 

For their operation, each motor must have a track-binding site and 
another site that binds and hydrolyzes ATP (so-called ATPase site). 
Both these sites are located, for example, in the {\it head} 
domain of myosins and kinesins both of which walk on their heads! 
The motor-binding sites on the tracks are equispaced; the actual 
step size of a motor can be, in principle, an integral multiple of 
the minimum step size which is the separation between two neighboring 
motor-binding sites on the corresponding track.

\subsubsection{Porters}

Some linear motors are {\it cargo transporters} \cite{mallik06}; 
however, the size of the cargoes are usually much larger than the 
motor itself! It is desirable that such a motor 
``walks'' for a significant distance on its track carrying the 
cargo; for obvious reasons, such motors are referred to as {\it 
porters} \cite{leibler93}. Kinesin and dyneins attached simultaneously 
to the same ``hard'' cargo can get engaged in a tug-of-war leading 
to a bidirectional movement of the cargo \cite{gross04,welte04}. 
In regions of overlap between MT and F-actin filaments, a large 
cargo may be hauled simultaneously by kinesins and myosins which,  
however, walk on their respective tracks. 
Alternatively, along its journey route, a cargo may be transferred 
from the MT-based transport network, which dominate at the cell center, 
to the F-actin based network that covers most of the cell periphery 
\cite{maniak03}.
A ``soft'' cargo pulled by many kinesins can get elongated into a 
tube \cite{leduc10}.

\subsubsection{Machines for chipping filamentous tracks}

A MT {\it depolymerase} is a kinesin motor that chips away its own track 
from one end \cite{howard07}. Members of the kinesin-13 family can reach 
either end of the MT diffusively (without ATP hydrolysis) and, then, 
start chipping the track from the end where it reaches. In contrast, 
members of the kinesin-8 family walk towards the plus end of the MT track 
hydrolyzing ATP and after reaching that end starts chipping it from there 
\cite{klein05,govindan08}. 
Chipping by both families of depolymerase kinesins are energized by ATP 
hydrolysis.

\subsubsection{Contractility: motor-filament crossbridge and collective dynamics of sliders and rowers}

Some motors are capable of sliding two different filaments with respect 
to each other by stepping simulatenously on these two filaments 
\cite{zemel09}. Some {\it sliders} work in groups and each detaches from 
the filament after every single stroke; these are often referred to as 
{\it rowers} because of the analogy with rowing with oars \cite{leibler93}. 
{\it Contractility}, rather than motility, at the subcellular and cellular 
level are driven by the sliders and rowers \cite{sellers04,squire05}. 
Some examples of this category are listed in the table \ref{table-rowslide}; 
the details can be found in the cited review articles.

\begin{widetext}
\begin{table}
\begin{tabular}{|c|c|c|} \hline
Motor & Sliding filaments  & Function (example) \\\hline
Myosin & ``Thin filaments'' (F-actin) of muscle fibers & Muscle contraction \cite{jontes95}\\ \hline
Myosin & ``Stress fibers'' (F-actin) of non-muscle cells & Cell contraction \cite{pellegrin07}\\ \hline
Myosin & Cytokinetic ``contractile ring'' (F-actin) in eukaryotes & Cell division \cite{zumdieck07} \\ \hline
Kinesin & Interpolar microtubules in mitotic spindle & Mitosis \cite{karsenti01,scholey10}\\ \hline
Dynein & Microtubules of axoneme & Beating of eukaryotic flagella \cite{satir10,lindemann10}\\ \hline
Dynein & Microtubules of megakaryocytes & Blood platelet formation \cite{italiano07} \\ \hline
\end{tabular}
\caption{Few example of cytoskeletal rowers and sliders as well as their biological functions.
 }
\label{table-rowslide}
\end{table}
\end{widetext}

\begin{widetext}
\begin{table}
\begin{tabular}{|c|c|c|} \hline
Polymer  & mode of force generation & Function (example) \\\hline
MT & polymerizing pistion-like & organizing cell interior \cite{tolic08}\\\hline
F-actin & polymerization & cell motility \cite{carlier10}  \\\hline
FtsZ & polymerization & bacterial cytokinesis \cite{erickson10}  \\\hline
MSP & polymerization & motility of nematode sperm cells \cite{wolgemuth05b} \\\hline
Type-IV pili & polymerization  & bacterial motility \cite{nudleman04}  \\\hline
MT & de-polymerization & Eukaryotic chromosome segregation \cite{mcintosh10}\\\hline
spasmin & spring-like & vorticellid spasmoneme \cite{mahadevan00} \\\hline
Coiled actin & spring-like & egg fertilization by sperm cell of the horse-shoe crab {\it Limulus polyphemus}  \cite{mahadevan11}  \\\hline
\end{tabular}
\caption{Force generation by polymerizing/depolymerizing, coiling/uncoiling filaments: pistons, hooks and springs.
 }
\label{table-piston}
\end{table}
\end{widetext}

\subsubsection{Push and pull of cytoskeletal filaments: nano-pistons and nano-springs}

Elongation of filamentous biopolymers that presses against a light object 
(e.g., a membrane) can result in a ``push'' \cite{mogilner03a}. 
Similarly, a depolymerizing tubular filament can ``pull'' a light  
ring-like object by inserting its hook-like outwardly curled 
depolymerizing tip into the ring \cite{mcintosh10}. 
A flexible filament, upon compression by input energy, can store energy 
that can perform mechanical work when the filament springs back to its 
original relaxed shape \cite{mahadevan00}. Some typical examples are 
given in table\ref{table-piston}.

\subsection{Machines for synthesis, manipulation and degradation of macromolecules of life}

\subsubsection{Membrane-associated machines for macromolecule translocation: exporters, importers and packers}

In many situations, the motor remains immobile and pulls a macromolecule; 
the latter are often called {\it translocase}. Some translocases 
{\it export} (or, {\it import}) either a protein \cite{schatz96} or 
a nucleic acid strand \cite{burton10,stewart10} across the plasma membrane 
of the cell or, in case of eukaryotes, across internal membranes. A list 
is provided in table \ref{table-memtranslocase}. 

The genome of many viruses are packaged into a pre-fabricated empty 
container, called {\it viral capsid}, by a powerful motor attached to 
the entrance of the capsid. As the capsid gets filled, The pressure 
inside the capsid increases which opposes further filling \cite{guo07,yu10}. 
The effective force, which opposes packaging, gets contributions from 
three sources: (a) bending of stiff DNA molecule inside the capsid;
(b) strong electrostatic repulsion between the negatively charged
strands of the DNA; (c) loss of entropy caused by the packaging 
\cite{petrov07a}.
The packaging motor has to be powerful enough to overcome such a 
high pressure.

\begin{table}
\begin{tabular}{|c|c|} \hline
Membrane & Polymer \\\hline
Nuclear envelope & RNA/Protein \cite{tu11,stewart10}  \\\hline
Membrane of endoplasmic reticulum & Protein \cite{rapoport08} \\\hline
Membranes of mitochondria/chloroplasts & Protein \cite{neupert07,aronson08}   \\\hline
Membrane of peroxisome & Protein \cite{gould02}   \\\hline
\end{tabular}
\caption{Membrane-bound translocases.
 }
\label{table-memtranslocase}
\end{table}

\subsubsection{Machines for degrading macromolecules of life}

Restriction-modification (RM) enzyme defend bacterial hosts against 
bacteriophage infection by cleaving the phage genome while the DNA 
of the host bacteria are not cleaved.
{\it Exosome} and {\it proteasome} are machines that shred RNA and 
proteins into their basic subunits, namely, nucleotides and amino 
acids, respectively. Similarly, there are machines for degrading 
polysachharides, e.g., cellulosome (a cellulose degrading machine), 
starch degrading enzymes, chitin degrading enzyme (chitinase), etc. 
These machines are listed in table \ref{table-degradosomes}. 

\begin{table}
\begin{tabular}{|c|c|} \hline
Polymer  & Examples of Machines \\\hline
DNA (polynucleotide) & RM enzyme \cite{pingoud05}  \\\hline
RNA (polynucleotide) & Exosome \cite{buttner06} \\\hline
Protein (polypeptide) & Proteasome \cite{lorentzen06}   \\\hline
Cellulose (polysachharide) & Cellulosome \cite{bayer04}  \\\hline
Starch (polysachharide) & Starch degrad. enzyme \cite{smith05a}  \\\hline
Chitin (polysachharide) & Chitinase \cite{chuan06}  \\\hline
\end{tabular}
\caption{Machines for degradation of macromolecules of life.
 }
\label{table-degradosomes}
\end{table}

\subsubsection{Machines for template-dictated polymerization}

Two classes of biopolymers, namely, polynucleotides and polypeptides 
perform wide range of important functions in a living cell. DNA and RNA 
are examples of polynucleotides while proteins are polypeptides. 
Both polynucleotides and polypeptides are made from a limited number of
different species of monomeric building blocks, namely, nucleotides and 
amino acids,respectively. The sequence of the monomeric subunits to be used for 
synthesis of each of these are dictated by that of the corresponding template. 
These polymers are elongated, step-by-step, during their birth by
successive addition of monomers, one at a time. The template itself 
also serves as the track for the polymerizer machine that  
takes chemical energy as input to polymerize the biopolymer as well as 
for its own forward movement. Therefore, these machines are also referred 
to as motors.

Depending on the nature of the template and product nucleic acid strands, 
polymerases can be classified as DNA-dependent DNA polymerase (DdDP), 
DNA-dependent RNA polymerase (DdRP), etc.  as listed in the table 
\ref{table-polymerase}.
\begin{table}
\begin{tabular}{|c|c|c|c|} \hline
Machine & Template & Product & Function\\ \hline
DdDP & DNA & DNA & DNA replication \cite{kornberg92} \\ \hline
DdRP & DNA & RNA & Transcription \cite{bai06,herbert08,nobelRNAP}\\ \hline
RdDP & RNA & DNA & Reverse transcription \cite{herschhorn10}\\ \hline
RdRP & RNA & RNA & RNA replication \cite{barr10} \\ \hline
Ribosome & mRNA & Protein & Translation \cite{spirin98,frank10,nobelRIBO}\\ \hline
\end{tabular}
\caption{Types of polymerizing machines, the templates they use
and the corresponding product of polymerization.
 }
\label{table-polymerase}
\end{table}

\subsubsection{Unwrappers and unzippers of packaged DNA: chromatin 
remodellers and Helicases}

In an eukaryotic cell DNA is packaged in a hierarchical structure 
called {\it chromatin}. In order to use a single strand of the 
DNA as a template for transcription or replication, it has to be 
unpackaged either locally or globally. ATP-dependent chromatin 
remodelers \cite{clapier09} 
are motors that perform this unpackaging. However, 
only one of the strands of the unpackaged duplex DNA serves as a 
template; the duplex DNA is {\it unzipped} by a DNA helicase 
\cite{pyle08,garai08}.
Similarly, a RNA helicase unwinds a RNA secondary structure.  
During DNA replication, a helicase moves ahead of the polymerase,  
like a mine sweeper, unzipping the duplex DNA and dislodging other 
DNA-bound proteins. However, the transcriptional and translational 
machineries do not need assistance of any helicase because these 
are capable of unzipping DNA and unwinding RNA, respectively, on 
their own. A helicase can be monomeric, or dimeric or hexameric.

\subsection{Rotary motors}

Rotary molecular motors are, at least superficially, very similar to 
the motor of a hair dryer. Two rotary motors have been studied most 
extensively. (i) A rotary motor embedded in the membrane of bacteria 
drive the bacterial flagella \cite{berg04,sowa08} 
which, the bacteria use for their swimming in aqueous media. 
(ii) A rotary motor, called ATP synthase \cite{oster02a,ballmoos08}, 
is embedded on the membrane of mitochondria, the powerhouses of a cell.
A {\bf synthase} drives a chemical reaction, typically the synthesis of 
some product; the ATP synthase produces ATP, the ``energy currency'' 
of the cell, from ADP.

\subsection{Processivity, duty ratio, stall force and dwell time}

One can define {\it processivity} of a motor in three different ways:\\
(i) Average number of {\it chemical cycles} in between attachment and
the next detachment from the track;\\
(ii) {\it mean time} of a single run, i.e., in between an attachment 
and the next detachment of the motor from the track;\\
(iii) {\it mean distance} spanned by the motor on the track in
a single run.\\
Since the definitions (ii) and (iii) are directly accessible to experiments, 
these are more useful than the definition (i). Another related, but 
distinct, concept is that of {\it duty ratio} which is defined as the 
average fraction of the time that each head spends remaining attached to 
its track during one cycle.

To translocate processively, a motor may utilize one of the three 
following strategies:\\
{\it Strategy I}: the motor may have more than one track-binding domain
(oligomeric structure can give rise to such a possibility quite
naturally). Most of the cytoskeletal motors like conventional two-headed
kinesin use such a strategy \cite{valentine07}. One of the track-binding 
sites remains bound to the track while the other searches for its next 
binding site. \\
{\it Strategy II}: A motor may possess non-motor extra domains or some 
accessory protein(s) bound to it which can interact with the track even 
when none of the motor domains of the motor is attached to the track. 
Dynein seems to exploit dynactin \cite{kardon09} to enhance its own 
processivity.\\
{\it Strategy III}: it can use a ``clamp-like'' device to remain attached 
to the track; opening of the clamp will be required before the motor
detaches from the track. For example, DdDP utilizes this strategy 
\cite{indiani06}.

An external force directed opposite to the natural direction of walk of 
a motor is called a {\it load} force. Average velocity of a motor decreases 
with increasing load force. The magnitude of the load force at which 
the average velocity of the motor vanishes, is called the {\it stall 
force}. The {\it force-velocity relation} is one of the most fundamental 
characteristic properties of a motor. Its status in biophysics is 
comparable, for example, to that of the I-V characteristics of a device 
in semiconductor physics.

Two motors with identical average velocities may exhibits widely
different types of fluctuations. Therefore, as we'll show in detail, 
a deeper understanding of the operational mechanism of a motor can be 
gained from the distributions of their ``dwells'' at the successive 
spatial positions on the track.

\subsection{Fundamental general questions}

(i) {\it Single-molecule mechanism} \cite{block07}: How do the interactions 
among the component structural units of an individual motor, 
motor-track interactions, motor-ligand (fuel) interactions and 
the mechano-chemical kinetics of the system determine, for example,  
(a) the directionality, (b) processivity, (c) dwell-time distribution, 
(d) force-velocity relation, and (e) efficiency of transduction?
Does a given dimeric motor walk hand-over-hand or crawl like an 
inchworm, and why? Do the ATPase domains of a hexameric motor 
``fire'' (a) in series, or (b) in parallel, or, (c) in random 
sequence? 

(ii) {\it Multi-motor coordination in a ``workshop''}: The motors 
do not work in isolation {\it in-vivo}. What are the mechanisms  
and consequences of the spatio-temporal coordination of the motors?  
For example, a single mitotic spindle consists of many polymerizing and 
depolymerizing MTs, several types of cytoskeletal motors, including 
depolymerases \cite{scholey10}. 
A replisome, the workshop for DNA replication, has to coordinate 
the operations of clamps and clamp loaders with those of primases, 
polymerases, etc. \cite{kornberg92}. 
Similarly, a ribosome is a mobile platform on which the operations 
of several devices have to be coordinated properly during protein 
synthesis \cite{spirin98}. 
Well known co-directional as well as head-on collisions of polymerases 
and those of ribosomes can create traffic jam under some circumstances 
whereas a collision can restart stalled traffic in other circumstances   
\cite{pomerantz10}. 

\section{Motoring in a viscous fluid: from Newton to Langevin}

Force is one of the most fundamental quantities in physics. As we'll 
argue in this section, some of the forces which dominate the dynamics 
of molecular machines have negligible effect on macroscopic machines.

\subsection{Newton's equation: deterministic dynamics} 
 
For simplicity, let us first consider a hypothetical scenario where 
neither the tracks nor the fuel molecules are present in the aqueous 
medium in which there is a free (i.e., not bound to any other molecule) 
motor protein. Suppose the medium consists of only $N$ ``particle-like'' 
molecules and the center of mass of the motor protein is also represented 
by a ``particle''. Then, the exact trajectories of all these $N+1$ 
``particles'' can be obtained by solving the corresponding coupled 
{\it Newton's} equations that describe the dynamics of the $N+1$ particles. 
In reality, this approach is impractical because, even with the largest 
and fastest computers available at present, we cannot get the trajectories  
over time intervals of the order of $1$s which is relevant for majority 
of the motor proteins as long as $N$ is large ($N \sim 10^{23}$).

\subsection{Langevin equation: stochastic dynamics} 
 
A more pragmatic approach would be to solve only the equations of 
motion for the motor protein by treating the $N$ particles of the 
medium as merely the constituents of a reservoir. In other 
words, one monitors the motion in a 6-dimensional subspace of the full 
$6(N+1)$-dimensional phase space of the system. The price one has to 
pay for this simplification is that instead of the original Newton's 
equation for the motor protein, one has to now solve a {\it Langevin 
equation} that describes the ``Brownian'' motion of the motor protein. 
Since we are soon going to extend the discussion in the presence of 
filamentous tracks which are essentially one-dimensional, we write 
the Langevin equation for the ``Brownian'' motion of the motor 
protein in one-dimension $m(d^2x/dt^2) = F_{d} + \xi$ 
where $m$ is the mass of the protein,  $F_d = - \gamma v$ is the 
viscous drag and and $\xi$ is the random Brownian force. Note 
that the Langevin equation is neither deterministic nor symmetric 
under time-reversal ($t \to -t$).

\subsubsection{Motoring in a ``sticky'' medium: Purcell's idea} 
 
Using the Stokes formula $\gamma = 6 \pi \eta r$ for a globular protein 
of radius $r \sim 10 nm$ moving with an average velocity $v \sim 1 m/sec$
(corresponding to $1 nm/ns$) in the aqueous medium of viscosity
$\eta \sim 10^{-3} Pas$ we get an estimate $F_d \sim 200 pN$.
Surprisingly, this viscous drag force is about $200$ times larger
than the elastic force it experiences when stretched  by $1 nm$ !
Consequently, the {\it Reynold's number}, which is the ratio of 
the inertial and viscous forces, is at most of the order $O(10^{-2})$. 
The Reynold's number will be of the order of $10^{-2}$ if you ever 
try (not recommended) to swim in honey!

Now let us supply the tracks and fuel molecules to the motor protein 
in the same medium. The Langevin equation will now include additional 
terms $F_{ext}$ which represents the net force arising from the 
interaction of the motor protein with its track and the ligand; 
the ligand could be the fuel molecule or the molecules produced by 
its ``burning'' (e.g., ATP or ADP). Besides, since the dynamics of 
motor proteins is expected to be dominated by hydrodynamics at low 
Reynold's number \cite{purcell77}, one generally drops the inertial 
term. In this ``overdamped'' regime
\begin{equation}
\gamma v = F_{ext} + \xi 
\label{eq-langevin}
\end{equation}

\subsubsection{Motoring under random impacts from surroundings: Brownian force} 
 
The energy released by the hydrolysis of a single ATP molecule is about
$10^{-21} J$. Interestingly, the mean thermal energy $k_BT$ associated
with a molecule at a temperature of the order of $T \sim 100 K$, is
also $k_B T \sim 10^{-21} J$. Moreover, equating this thermal energy
with the work done by the thermal force $F_t$ in causing a displacement
of $1 nm$ we get $F_t \sim 1 pN$. This is comparable to the elastic
force experienced by a typical motor protein when stretched by $1$nm.
Thus, a motor protein that gets bombarded from all sides by random
thermal forces is similar to a tiny creature is a very strong hurricane!
Therefore, in contrast to the deterministic dynamics of the macroscopic
machines, the dynamics of molecular motors is stochastic (i.e.,
probabilistic).

Already in the first half of the twentieth century D'Arcy Thompson, 
father of modern bio-mechanics, realized the importance of {\it viscous 
drag} and {\it Brownian forces} at the cellular (and subcellular) level. 
He pointed out that \cite{darcy63} that in this microscopic world 
``{\it gravitation is forgotten, and the viscosity of the liquid,..., 
the molecular shocks of the Brownian movement, .... the 
electric charges of the ionized medium, make up the physical environment 
... predominant factors are no longer those of our scale}''.

\section{Energy transduction by molecular machines: From Carnot to Feynman}

Molecular motors generate force by transducing energy. Interestingly, 
one of the two mechanisms of energy transduction that I describe here, 
does not have any analogous macroscopic counterpart. I explain in 
considerable detail why thermodynamic formalisms, which have been 
successfully utilized to calculate the common performance characteristics 
of macroscopic machines, are inadequate for natural nano-machines. 

Molecular motors are made of {\it soft matter} whereas macroscopic 
motors are normally made of hard matter to withstand wear and tear. 
Nature seems to exploit the high deformability of molecular motors 
for its biological function. But, this difference is minor compared 
to the others and will not be elaborated further here.

\subsection{Molecular machines are run by isothermal engines} 

This is in sharp contrast to the macroscopic thermal engines which 
require at least two thermal reservoirs at different temperatures 
and which convert part of the input heat energy into mechanical work. 

Why can't a molecular machine work as a heat engine? In order to examine 
the viability of a intracellular heat engine, let us create a temperature 
gradient on a length scale ${\ell} \sim 10$nm. An elementary analysis 
of heat diffusion equation is adequate to show that a temperature gradient 
on a length scale ${\ell}$ relaxes within a time interval 
$\tau_{temp} \sim c_{h} {\ell}^{2}/\kappa$ 
where $c_h$ is the specific heat per unit volume and $\kappa$ is the 
thermal conductivity.  
Using the typical characteristic values of $c_{h}$ and $\kappa$ for 
water, one finds \cite{julicher06} 
$\tau_{temp} \simeq 10^{-6} - 10^{-8}$s. Thus, temperature gradients
cannot be maintained for the entire duration of even one single cycle
of a cyclic molecular machine which is, typically, few orders of
magnitude longer than $\tau_{temp}$. In other words, for all practical
purposes, natural nano-machines operate {\it isothermally}. Because
of the condition $T = constant$, the molecular machines operate as
{\it free energy transducers}.

\subsection{Inadequacy of equilibrium thermodynamics} 

A majority of the molecular motors are chemo-mechanical machines 
for which input and output are chemical energy and mechanical work, 
respectively. Recall that for Carnot's thermo-mechanical machine, the 
thermodynamic efficiency is $\eta_{eq-th} = 1-(T_{L}/T_{H})$ where 
$T_{H}$ and $T_{L}$ are the temperatures of the two reservoirs at high 
and low temperatures, respectively. Similarly, for an isothermal 
chemo-mechanical engine, the thermal reservoirs would be replaced by 
chemical reservoirs at fixed chemical potentials $\mu_H$ and $\mu_L$ 
($\mu_{H} > \mu_{L}$). Consequently, the heat flow of the thermal 
engine would be replaced by particle flow in the chemical engine. Just 
as the difference of heat input and output in the heat engine is 
converted to mechanical work, the difference $\mu_H-\mu_L$ would be 
converted into output work in the chemical engine. Obviously, 
the corresponding thermodynamic efficiency would be \cite{sasaki05} 
$\eta_{eq-ch} = 1 - (\mu_L/\mu_H)$. Note that, because of the 
quasi-static nature of each step in the equilibrium thermodynamic 
theory of these cyclic engines, each cycle takes infinite time. 
Naturally, the power output ${\cal P}_{out} = 0$ for both the thermal 
and chemical Carnot engines. 

Can we apply the theory of such isothermal Carnot engines to a 
chemo-mechanical molecular machine by identifying, for example, 
$\mu_H$ and $\mu_L$ ($\mu_{H} > \mu_{L}$)  
as the chemical potentials of ATP and ADP, respectively?  
The answer is: NO. The cycle time of the real molecular machines is 
{\it finite} and their power output is non-zero. 
Moreover, motors are examples of open systems \cite{bertalanffy50} 
which continue to run in repetitive cycles as long as energy is 
pumped in. Such a system cannot be in thermodynamics equilibrium. 

\subsection{Inadequacy of endo-reversible thermodynamics} 

For macroscopic engines with {\it finite cycle time}, the characteristics 
of performance are often reliably calculated within the theoretical 
framework of endo-reversible thermodynamics \cite{berry00a}. In this case,  
the ratio of the {\it rates} of output and input energies defines the  
efficiency of transduction. The rate of output work is called power output 
of the engine.

The formalism of endo-reversible thermodyamics for {\it heat} engines 
is based on the {\it assumption} that the working substance of the 
engine is coupled to the two {\it thermal} reservoirs by heat conductors 
of {\it finite} thermal conductivity (non-zero thermal resistance). 
Entire dissipation is assumed to take place in the irreversible process 
of heat conduction through these thermal resistors whereas all the 
processes in the working material of the engine are assumed to be fully 
reversible (i.e., no entropy is generated internally). In this scenario, 
the corresponding thermal conductances in the Carnot engine are infinite.  

Efficiency at maximum power output $\eta({\cal P}_{max})$, rather than 
maximum efficiency itself, is the most appropriate quantitative 
measure of the performance of such engines. The upper-bound of 
$\eta({\cal P}_{max})$ is given by the Curzon-Ahlborn expression 
\cite{curzon75} 
$\eta_{CA}({\cal P}_{max}) = 1 - (T_{L}/T_{H})^{1/2}$.  
Similarly, within the framework of the endoreversible thermodynamics, 
a phenomenological theory for {\it chemical} engines can also be 
formulated if the cycle time is {\it finite} \cite{gordon93}. 
Can't we use this approach for molecular machines which are also 
chemical engines with finite cycle times? 

Recall that endo-reversible thermodynamics is based on the assumption 
that dissipation takes place only in the process of transfer of matter 
between reservoirs and the working substance whereas processes that 
the working substance goes through in each cycle are perfectly reversible 
(and, therefore, non-dissipative). This assumption is valid if the 
relaxations in the working substance of the engine are very rapid compared to 
the processes involving the coupling of the working substance with the 
reservoirs. The validity of this assumption requires clear separation 
of time scales of relaxation in the working substance and in the working 
substance-reservoir coupling. But, such separation of time scales does 
not hold in molecular machines which consist of macromolecules.

\subsection{Domain of non-equilibrium statistical mechanics}

In general, molecular motors operate far beyond the linear response 
regime and, therefore, the formalism of non-equilibrium thermodynamics 
\cite{katchalsky67} for coupled mechano-chemical processes is not 
applicable. Moreover, thermodynamic formalisms developed for 
macroscopically large systems do not account for the spontaneous 
fluctuations. On the other hand, because of the small size of a 
molecular motor and because of the low concentrations of the molecules 
involved in its operation, fluctuations of positions, conformations 
as well as the cycle times are intrinsic features of their kinetics. 
Therefore, one has to use the more sophisticated toolbox of stochastic 
processes and non-equilibrium statistical mechanics for theoretical 
treatment of molecular machines. Interestingly, as we argue below, 
noise need not be a nuisance for a motor; instead, a motor can move 
forward by gainfully exploiting this noise!

\subsection{Defining efficiency: from Carnot to Stokes}

The performance of {\it macroscopic} motors are characterized by a 
combination of its efficiency, power output, maximum force or torque 
that it can generate. Just like the performance of their macroscopic 
counterparts with finite cycle time, that of molecular motors 
\cite{schmiedl08} have also been characterized in terms of efficiency 
at maximum power, rather than maximum efficiency. However, the 
efficiency of molecular motors can be defined in several different 
ways \cite{linke05}. 

The efficiency of a motor, with finite cycle time, is generally 
defined by  
\begin{equation}
\eta = {\cal P}_{out}/{\cal P}_{in}
\label{eq-eff}
\end{equation} 
where ${\cal P}_{in}$ and ${\cal P}_{out}$ are the input and output 
powers, respectively. The usual definition of {\it thermodynamic 
efficiency} $\eta_{T}$ is based on the assumption that, like its 
macroscopic counterpart, a molecular motor has an output power 
\cite{parmeggiani99} 
\begin{equation}
{\cal P}_{out} = - F_{ext} V. 
\label{eq-thermoout}
\end{equation} 
where $F_{ext}$ is the externally applied opposing (load) force. 
Although this definition is unambiguous, it is unsatisfactory for 
practical use in characterizing the performance of molecular 
motors. As explained earlier, a molecular motor has to work against 
the omnipresent viscous drag in the intracelluar medium even when 
no other force opposes its movement (i.e., even if $F_{ext}=0$). 

A generalized efficiency $\eta_{G}$ is also represented by the 
same expression (\ref{eq-eff}) where, instead of (\ref{eq-thermoout}), 
the output power is assumed to be \cite{derenyi99a} 
\begin{equation}
{\cal P}_{out} = F_{ext} V + \gamma V^2.
\label{eq-genout}
\end{equation} 
This definition treats the load force and viscous drag on equal footing. 
In contrast, the ``Stokes efficiency'' $\eta_{S}$ for a molecular motor 
driven by a chemical reaction is defined as \cite{wang02b} 
\begin{equation}
\eta_{S} = \frac{\gamma V^2}{(\Delta G)\langle r \rangle + F_{ext} V}
\end{equation}
where $\langle r \rangle$ is the average rate of the chemical reaction 
and $\Delta G$ is the chemical free energy consumed in each reaction 
cycle. This efficiency is named after Stokes because the viscous drag 
is calculated from Stokes law. 

As we show in the next subsection, the directional movement of some 
motors arises from the rectification of random thermal noise. For 
such motors, an alternative measure of performance is the {\it 
rectification efficiency} \cite{suzuki03}. 
Thus, different definitions of the efficiency of a molecular motor may 
arise from different interpretations of their tasks or may characterize 
distinct aspects of their operation. Nevertheless, for any definition 
of efficiency, it is essential to ensure that its allowed values lie 
between $0$ and $1$.

\subsection{Noisy power stroke and Brownian ratchet}

If the input energy directly causes a conformational change of the  
protein machinery which manifests itself as a mechanical stroke of 
the motor, the operation of the motor is said to be driven by a 
``power stroke'' mechanism. This is also the mechanism used by all 
man made macroscopic machines. However, in case of molecular motors, 
the power stroke is always ``noisy'' because of the Brownian forces 
acting on it.

Let us contrast this with the following alternative scenario: 
suppose, the machine exhibits both ``forward'' and ``backward'' 
movements because of spontaneous thermal fluctuations. If now energy 
input is utilized to prevent ``backward'' movements, but allow the 
``forward'' movements, the system will exhibit directed, albeit noisy, 
movement in the ``forward'' direction. Note that the forward movements in 
this case are caused directly by the spontaneous thermal fluctuations, 
the input energy rectifies the ``backward'' movements. This alternative 
scenario is called the Brownian ratchet mechanism 
\cite{julicher97a,astumian07}. 
The concept of Brownian ratchet \cite{reimann02a} 
was popularized by Richard Feynman with his ratchet-and-pawl device 
\cite{feynmanbook}. 
However, in the context of molecular motors, the Brownin ratchet 
mechanism was proposed by several groups in the 1990s 
\cite{vale90,julicher97a}. 

Thus, in principle, there are two idealized scenarios for a transition 
from a conformation A to a conformation B- one is by a pure power stroke 
and the other by a purely Brownian ratchet mechanism \cite{howard06a}. 
However, for a real molecular motor, it is difficult to unambiguously 
distinguish between a power stroke and a Brownian ratchet \cite{wang02a}; 
the actual mechanism may be a combination of the two idealized extremes.

\noindent$\bullet${\bf Are Brownian motors Maxwell's demon?}

Brownian ratchet mechanism transduces random thermal energy of the 
reservoir into mechanical work. Does it imply that these motors 
are perpetual motors of the second kind that violate the second 
law of thermodynamics? In that case, does it 
have any similarity with the Maxwell's demon \cite{leff02}?
A short answer is: No, a Brownian ratchet is not a Maxwell's 
demon because it is an open system far from the state of equilibrium 
whereas the second law of thermodynamics is strictly valid only 
for systems in stable thermodynamic equilibrium. 

\subsection{Physical realizations of Brownian ratchet in molecular motors}

Is there any real biomolecular motor which can be regarded as a true 
physical realization of Brownian ratchets? The answer is an emphatic 
``yes'' and we list a few typical examples here. 

KIF1A, a single-headed kinesin, is an example of cytoskeletal 
motor whose operational mechanism can be interpreted as a Brownian 
ratchet \cite{okada00,nishinari05,greulich07}. 
The original acto-myosin crossbridge model suggested by Huxley 
\cite{ahuxley57} can be interpreted as Brownian ratchet \cite{cordova92} 
although Huxley himself did not use this terminology. 
Brownian ratchet can also account for the force generation by a 
polymerizing cytoskeletal filament \cite{peskin93b}. 

The crossing of a membrane during the export/import of a protein of 
length $L$ (amino acid monomers) takes place at a faster rate by the 
Brownian ratchet mechanism compared to that by pure translational 
diffusion \cite{simon92}. 
It has been claimed that the translocation of a mRNA molecule across 
the nuclear membrane of an eukaryotic cell takes place also by a 
similar Brownian ratchet mechanism through the nuclear pore complex 
\cite{stewart07a}. 
The elongation of a mRNA during transcription by a RNA polymerase 
can also be a physical realization of the Brownian ratchet mechanism 
\cite{barnahum05,guo06}. 
Translocation, one of the crucial steps of the mechano-chemical cycle 
of a ribosome is a physical realization of Brownian ratchet \cite{frank10}.

\section{Mathematical description of mechano-chemical kinetics: continuous landscapes vs. discrete networks} 
\label{sec-math}

We combine the fundamental principles of (stochastic) nano-mechanics and 
(stochastic) chemical kinetics to formulate the general theoretical 
framework for a quantitative description of the {\it mechano-chemistry}  
or {\it chemo-mechanics} of molecular motors. We mention a few alternative 
formalisms.

\subsection{\bf Motor kinetics as wandering in a time-independent mechano-chemical free-energy landscape} 

This approach is useful for an intuitive physical explanation of the 
coupled mechano-chemical kinetics of molecular motors. At least one 
of the independent coordinate axes of this landscape represents the 
position of the motor in real space (i.e., its ``mechanical coordinate'') 
while at least another represents its chemical state (.e., ``chemical 
coordinate''). Therefore, the minimum dimension of this ``land'' is 
$2$ and the free energy can be represented by the hight at each point 
on the ``land''. Both the position and chemical state variables are 
assumed to be continuous. The profiles of the free energy along both 
the position and chemical coordinates, obtained by taking appropriate 
cross sections of the free energy landscape, are periodic with the 
same periodicity in both the directions. But, the profile is tiled 
forward along the chemical direction so that the bottom of the minima 
are located deeper in the forward direction along the ``chemical'' 
coordinate; this tilt accounts for the lowering of free energy caused, 
for example, by ATP hydrolysis.

\subsection{\bf Motor kinetics as wandering in the time-dependent real-space potential landscape}

Consider those special situations where the chemical states of the motor 
are {\it long lived} and change in {\it fast discrete jumps} so that 
the position of the motor can continue to change without alteration in 
its chemical state, except during chemical transitions when position 
remains frozen. Thus, no mixed mechano-chemical transition is allowed 
in this scenario. 
In such situations, we can assume that the potential landscape in 
{\it real space} remains unchanged for a while, during which the 
wanderings of the motor in this landscape (i.e., the positional 
dynamics of the motor) is governed by the ``frozen'' spatial profile 
of the potential. The profile of the potential switches sequentially 
from one form to another and the sequence of $m$ profiles is repeated 
in each cycle although the switching times are random. Different 
profiles correspond to different motor-track interactions which is 
dependent on the nature of the ligand (if any) bound to the motor.

In this formulation we assume that the allowed positions of the motor 
on its track form a {\it continuum} $x$. The {\it discrete} index 
$\mu=1,2,...,M$ denote the $M$ distinct spatial profiles of the potential 
$V_{\mu}(x)$ that are postulated {\it apriori}. At any instant of time 
$t$, the state of the motor is given by $(x,\mu)$. 
In the overdamped regime, the time-evolution of the position $x(t)$ of 
the motor obeys the Langevin equation (\ref{eq-langevin}) with 
\cite{wang07a}
$F_{ext} = - dV_{\mu}(x)/dx + F_{load}$. 
The time-dependence of the profile of the potential $V_{\mu}(x)$ 
is governed by the master equation
\begin{eqnarray}
\frac{\partial P_{\mu}(x,t)}{\partial t} = \sum_{\mu'} P_{\mu'}(x,t) W_{\mu'\to \mu}(x) - \sum_{\mu'} P_{\mu}(x,t) W_{\mu \to \mu'}(x) \nonumber \\
\label{eq-chmaster}
\end{eqnarray}
where $P_{\mu}(x,t)$ is the probability that the motor protein 
``sees'' the landscape $V_{\mu}(x)$ at time $t$ and $W_{\mu \to \mu'}(x)$ 
is the transition probability per unit time for a transition from 
the chemical state $\mu$ to the chemical state $\mu'$ at position $x$.

Instead of the Langevin equation, an equivalent Fokker-Planck equation 
can be used to describe the wandering of the motor protein in the 
potential energy landscape $V_{\mu}(x)$. The advantage of this alternative 
formulation is that the Fokker-Planck equation can be combined naturally 
with the master equation that describes the time-evolution of the potential 
energy landscape. In this hybrid formulation, $P_{\mu}(x,t)$ represents 
the probability that, at time $t$, the motor is located at $x$, while 
``seeing'' the potential energy landscape $V_{\mu}(x)$. The equation 
governing the time dependence of $P_{\mu}(x,t)$
\begin{eqnarray}
\frac{\partial P_{\mu}(x,t)}{\partial t} &=& \frac{1}{\eta} \frac{\partial}{\partial x}\biggl[\{V'_{\mu}(x) - F\} P_{\mu}(x,t)\biggr] \nonumber \\
&+& \biggl(\frac{k_BT}{\eta}\biggr) \frac{\partial^2 P_{\mu}(x,t)}{\partial x^2} \nonumber \\
&+& \sum_{\mu'} P_{\mu'}(x,t) W_{\mu'\to \mu}(x) \nonumber \\
&-& \sum_{\mu'} P_{\mu}(x,t) W_{\mu\to \mu'}(x)
\label{eq-hybrid}
\end{eqnarray}
where $\eta$ is a phenomenological coefficient that captures the effect 
of viscous drag.
Note that there is no term in this equation which would correspond to 
a mixed mechano-chemical transition.

\subsection{\bf Motor kinetics as a jump process in a fully discrete mechano-chemical network}

In this formulation, both the positions and ``internal'' (or ``chemical'') 
states of the motors are assumed to be discrete.
Let $P_{\mu}(i,t)$ be the probability of finding the motor at the
discrete position labelled by $i$ and in the ``chemical''
state $\mu$ at time $t$. Then, the master equation for $P_{\mu}(i,t)$
is given by
\begin{eqnarray}
\frac{\partial P_{\mu}(i,t)}{\partial t}&=&[\sum_{j\neq i} P_{\mu}(j,t) k_{\mu}(j \to i) - \sum_{j \neq i} P_{\mu}(i,t) k_{\mu}(i \to j)] \nonumber \\ 
&+& [\sum_{\mu'} P_{\mu'}(i,t) W_{\mu'\to \mu}(i) - \sum_{\mu'} P_{\mu}(i,t) W_{\mu\to \mu'}(i)] \nonumber \\
&+& [\sum_{j\neq i} \sum_{\mu'} P_{\mu'}(j,t) \omega_{\mu'\to \mu}(j \to i) \nonumber \\
&-& \sum_{j\neq i} \sum_{\mu'} P_{\mu}(i,t) \omega_{\mu\to \mu'}(i \to j)]
\label{eq-fullmaster}
\end{eqnarray}
where the terms enclosed by the three different brackets $[.]$ correspond
to the purely mechanical, purely chemical and mechano-chemical transitions,
respectively.

As a concrete example, which will be used also for on several other 
occasions later in this review, consider a 2-state motor that, at 
any site $j$, can exist in one of the only two possible chemical 
states labelled by the symbols $1_j$ and $2_j$. We assume the 
mechano-chemical cycle of this motor to be 
\begin{equation}
1_{j} \mathop{\rightleftharpoons}^{\omega_{1}}_{\omega_{-1}} 2_{j} \mathop{\rightarrow}^{\omega_{2}} 1_{j+1}
\label{eq-MMlikemot}
\end{equation}
where the rates of the allowed transitions are shown explicitly 
above or below the corresponding arrow. Note that the transition 
$1_j \rightleftharpoons 2_{j}$ is purely chemical whereas the 
transition $2_{j} \rightarrow 1_{j+1}$ is a mixed mechano-chemical 
transition. The corresponding master equations are given by 
\begin{eqnarray}
\frac{dP_1(i)}{dt} = \omega_{2} P_2(i-1) + \omega_{-1} P_{2}(i) - \omega_{1} P_{1}(i) \nonumber \\
\frac{dP_2(i)}{dt} = \omega_{1} P_1(i) - \omega_{-1} P_{2}(i) - \omega_{2} P_{2}(i) \nonumber \\
\end{eqnarray} 
Suppose a load force $F_{ext}$ opposes the meacho-chemical transition 
$2_{j} \rightarrow 1_{j+1}$. The effect of the load force can be 
incorporated by assuming the load-dependence of the corresponding rate 
constant to be of the form 
$\omega_{2}(F_{ext}) = \omega_{2}(0) exp[-F_{ext}/(k_BT)]$ where 
$\omega_{2}(0)$ is the value of the rate constant in the absence 
of any external force.
We'll see some implications of these equations in several specific 
contexts later in this article.

\subsection{Balance conditions for mechano-chemical kinetics: cycles, and flux}

On a discrete mechano-chemical state space, each state is denoted by 
a {\it vertex} and the direct transition from one state (denoted by, 
say, the vertex $i$) to another (denoted by, say, the vertex $j$) is 
represented by a directed {\it edge} $|ij>$. The opposite transition 
from $j$ to $i$ is denoted by the directed edge $|ji>$.  
A {\it transition flux} can be defined along any edge of this diagram. 
The forward transition flux $J_{|ij>}$ from the vertex $i$ to the 
vertex $j$ is given by $W_{ji} P_{i}$ while the reverse flux, i.e., 
transition flux $J_{|ji>}$ from $j$ to $i$ is given by $W_{ij} P_{j}$. 
Therefore, the net transition flux in the direction {\it from} the 
vertex $i$ {\it to} the vertex $j$ is given by 
$J_{ji} = W_{ji} P_{i} - W_{ij} P_{j}$. 

A {\it cycle} in the kinetic diagram consists of at least three 
vertices. Each cycle $C_{\mu}$ can be decomposed into two directed 
cycles (or, {\it dicycles}) \cite{lipowsky08b} $C_{\mu}^{d}$ where 
$d=\pm$ corresponds the clockwise and counter-clockwise cycles. 
The net cycle flux $J(C_{\mu})$ in the cycle $C_{\mu}$, in the CW 
direction, is given by $J(C_{\mu}) = J(C_{\mu}^{+}) - J(C_{\mu}^{-})$. 

For each arbitrary dicycle $C_{\mu}^{d}$, let us define the {\it 
dicycle ratio} 
\begin{equation}
{\cal R}(C_{\mu}^{d}) = \Pi_{<ij>\epsilon C_{\mu}^{+d}} W_{ji}/\Pi_{<ij>\epsilon C_{\mu}^{-d}} W_{ij} = \Pi_{<ij>}^{\mu,d} (W_{ji}/W_{ij}).
\end{equation}
where the superscript $\mu,d$ on the product sign denote a product 
evaluated over the directed edges of the cycle. So, by definition, 
${\cal R}(C_{\mu}^{-}) = 1/{\cal R}(C_{\mu}^{+})$.

It has been proved rigorously \cite{vankampenbook} that, for detailed 
balance to hold, the necessary and sufficient condition to be 
satisfied by the transition probabilities is   
\begin{equation}
{\cal R}(C_{\mu}^{d}) = 1 ~~{\rm for ~all ~dicycles}~~ C_{\mu}^{d}
\end{equation}

For a non-equilibrium steady state (NESS), one can define 
\cite{lipowsky08b} the {\it dicycle frequency} 
$\Omega^{ss}(C_{\mu}^{d})$ which is the number of dicycles $C_{\mu}^{d}$ 
completed per unit time in the NESS of the system. Then, 
\begin{equation}
\Omega^{ss}(C_{\mu}^{+})/\Omega^{ss}(C_{\mu}^{-}) = \Pi_{<ij>}^{\mu,d} (W_{ji}/W_{ij}) = {\cal R}(C_{\mu}^{d})
\end{equation}
Clearly, ${\cal R}(C_{\mu}^{d}) \neq 1$, in general, for any NESS.

Detailed balance is believed to be a property of systems in equilibrium 
whereas the conditions under which molecular machines operate are far 
from equilibrium. Does it imply that detailed balance breaks down for  
molecular machines?  The correct answer this subtle question needs a 
careful analysis \cite{wang05,astumian05,thomas01a}.

If one naively {\it assumes} that the entire system returns to 
its original initial state at the end of each cycle one would 
get the erroneous result that the detailed balance breaks down. 
But, strictly speaking, the free energy of the full system gets 
lowered by $|\delta G|$ (e.g., because of the hydrolysis of ATP) 
in each cycle although the cyclic machine itself comes back to 
the same state. When the latter fact is incorporated correctly 
in the analysis \cite{wang05,astumian05,thomas01a}, 
one finds that detailed balance is not violated by molecular 
machines. This should not sound surprising- the transition rates 
``do not know'' whether or not the system has been driven out of 
equilibrium by pumping energy into it.

\section{Solving forward problem by stochastic process modeling: from model to data}

\begin{figure}[ht]
\begin{center}
(a)\\
\includegraphics[angle=-90,width=0.65\columnwidth]{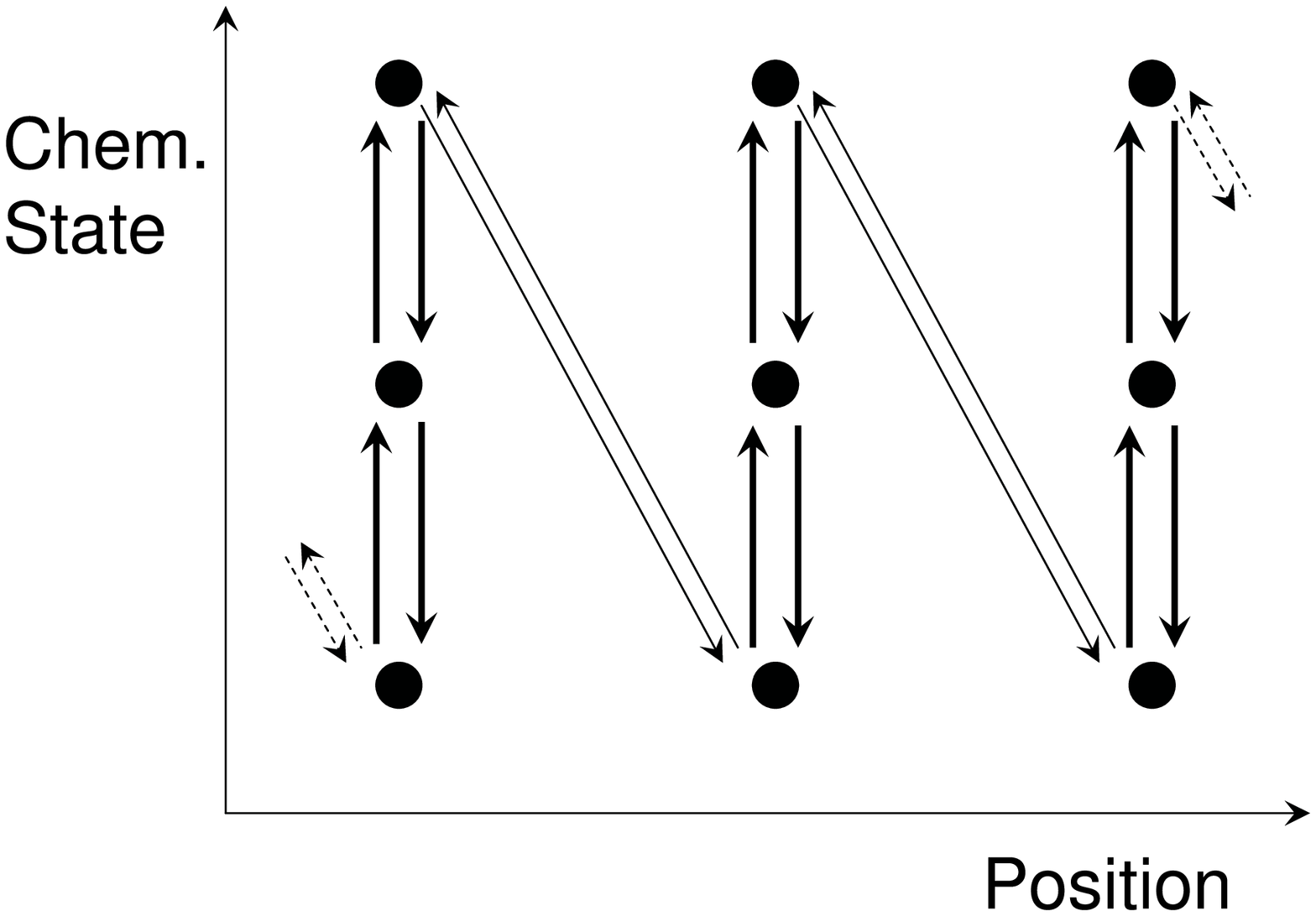}\\
(b)\\
\includegraphics[angle=-90,width=0.65\columnwidth]{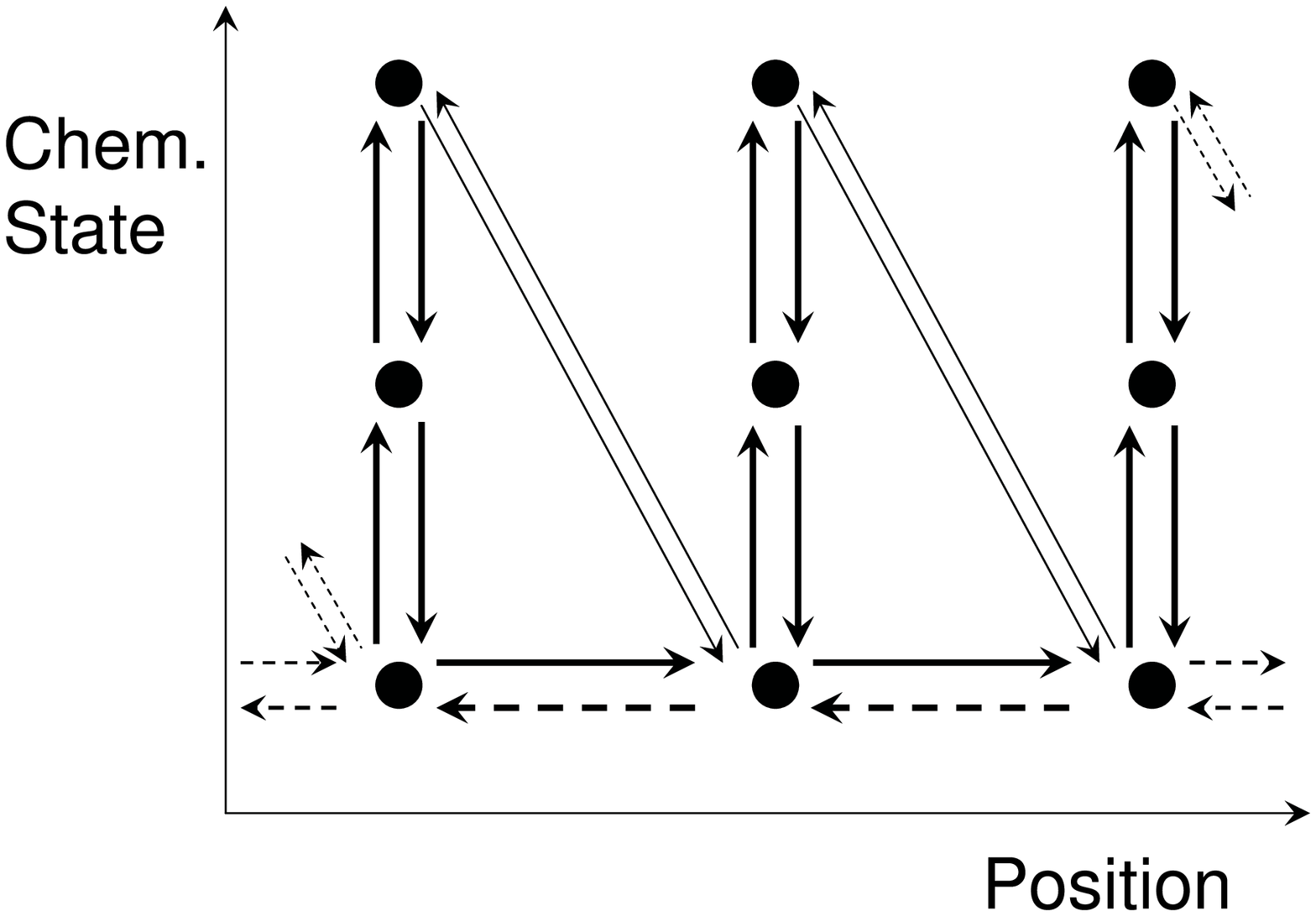}\\
(c)\\
\includegraphics[angle=-90,width=0.65\columnwidth]{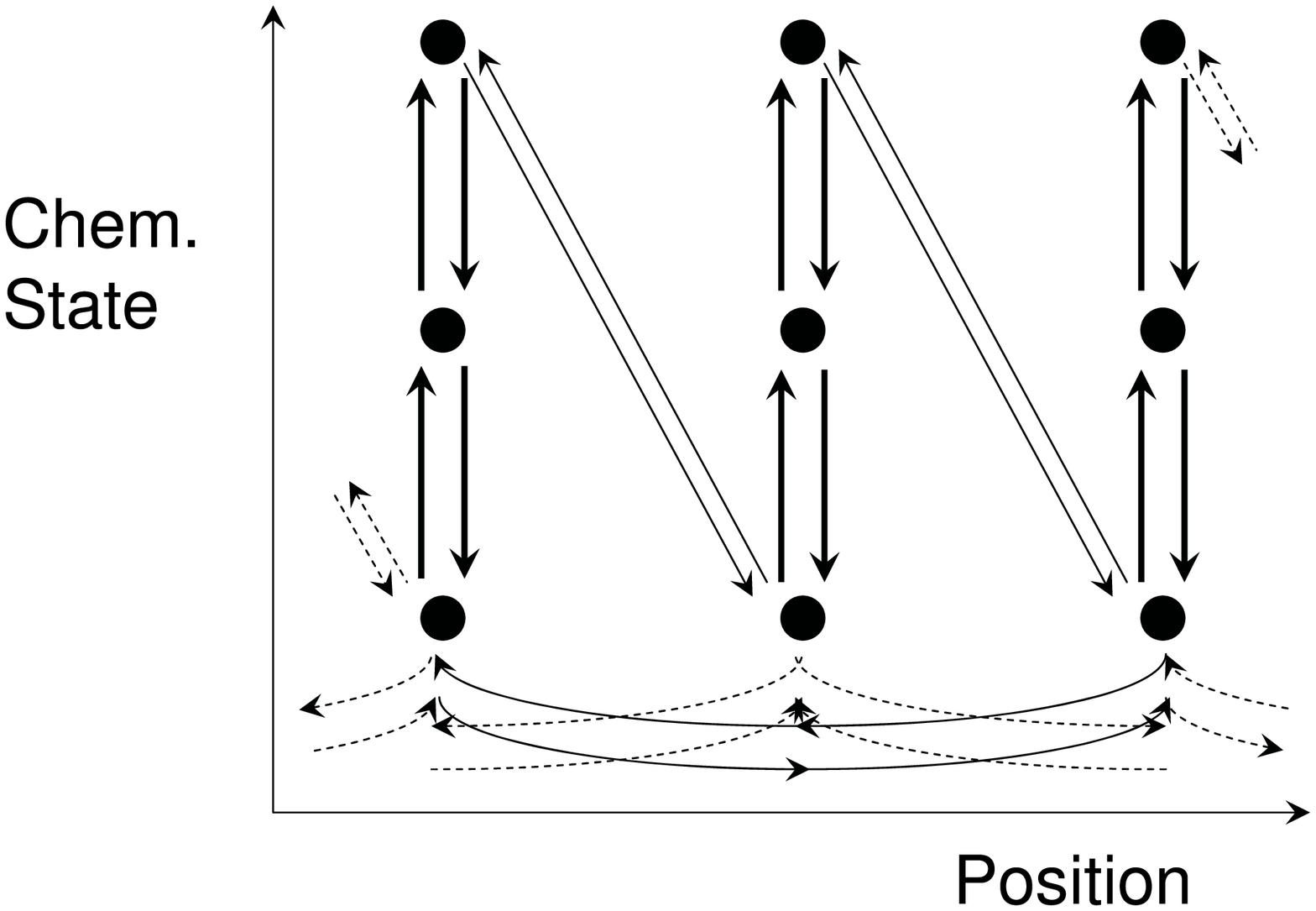}
\end{center}
\caption{Three examples of different types of networks of 
discrete mechano-chemical states. The bullets represent the 
distinct states and the arrows denote the allowed transitions 
between the two corresponding states. The scheme is (a) is 
unbrached whereas that in (b) has branched pathways connecting 
the same pair of states. The mechanical step size is unique 
in both (a) and (b) whereas steps of more than one size are 
possible in (c). (Adapted from fig.1 of ref.\cite{chemla08}).
}
\label{fig-DTD}
\end{figure}

\subsection{\bf Average speed and load-velocity relation }

For simplicity, let us consider the kinetic scheme shown in fig. 
(\ref{fig-DTD}(a)). In terms of the Fourier transform 
\begin{equation}
\bar{P}_{\mu}(k,t) = \sum_{j=-\infty}^{\infty} P_{\mu}(x_{j},t) e^{-ikx_j} 
\label{eq-FourierP}
\end{equation}
of $P_{\mu}(x_{j},t)$, the master equations can be written as a 
matrix equation 
\begin{equation}
\frac{\partial {\bf \bar{P}}(k,t)}{\partial t} = {\bf W}(k) {\bf \bar{P}}(k,t) 
\label{eq-masterq}
\end{equation} 
where ${\bf \bar{P}}$ is a column vector of $M$ components ($\mu=1,...M$) 
and ${\bf W}(k)$ is the transition matrix in the $k$-space (i.e., 
Fourier space). Summing over the hidden chemical states, we get 
\begin{equation}
\bar{P}(k,t) = \sum_{\mu=1}^M \bar{P}_{\mu}(k,t) = \sum_{\mu=1}^M \sum_{j=-\infty}^{\infty} P_{\mu}(x_j,t) e^{- i k x_j}. 
\label{eq-barPk}
\end{equation}
Taking derivatives of both sides of (\ref{eq-barPk}) with respect to 
$q$ we get \cite{koza99,koza00} 
\begin{eqnarray} 
i \biggl(\frac{\partial \bar{P}(k,t)}{\partial k}\biggr)_{k=0} &=& <x(t)> \nonumber \\
- \biggl(\frac{\partial^2 \bar{P}}{\partial k^2}\biggr)_{k=0} + \biggl(\frac{\partial P}{\partial k}\biggr)^2_{k=0} &=& <x^2(t)>-<x(t)>^2 \nonumber \\ 
\end{eqnarray}
Evaluating $\bar{P}(k,t)$, in principle, the stationary 
drift velocity (i.e., asymptotic mean velocity) $V$ and the 
corresponsding diffusion constant $D$ can be obtained from 
\begin{eqnarray} 
&&V = lim_{t \to \infty} i \frac{\partial}{\partial t}\biggl[\biggl( \frac{\partial \bar{P}(k,t)}{\partial k}\biggr)_{k=0}\biggr] \nonumber \\
&&D = lim_{t \to \infty} \frac{1}{2} \frac{\partial}{\partial t}\biggl[ \biggl(-\frac{\partial^2\bar{P}(k,t)}{\partial k^2}\biggr)_{k=0} + \biggl(\frac{\partial\bar{P}(k,t)}{\partial k}\biggr)^2_{k=0}\biggr] \nonumber \\
\end{eqnarray}

It may be tempting to attempt a direct utilization of the general form 
\begin{equation}
\bar{P}(k,t) = \sum_{\mu} B_{\mu} e^{\lambda_{\mu}(k) t} 
\label{eq-gensol}
\end{equation}
where the coefficients $B_{\mu}$ are fixed by the initial conditions and 
$\lambda_{\mu}(k)$ are the eigenvalues of ${\bf W}(k)$. 
However, for the practical implementation of this method analytically 
the main hurdle would be to get all the eigenvalues of ${\bf W}(k)$. 
Fortunately, only the smallest eigenvalue $\lambda_{min}$, which 
dominates $\bar{P}(k)$ 
in the limit $t \to \infty$, is required for evaluating $V$ and $D$ 
\cite{koza99,koza00}: 
\begin{eqnarray} 
V &=& i \biggl(\frac{\partial \lambda_{min}(k,t)}{\partial k}\biggr)_{k=0} \nonumber \\
D &=& - \frac{1}{2} \biggl(\frac{\partial^2\lambda_{min}(k,t)}{\partial k^2}\biggr)_{k=0} 
\label{eq-lambda}
\end{eqnarray}
Even the forms (\ref{eq-lambda}) are not convenient for evaluating 
$V$ and $D$. Most convenient approach is based on the characteristic 
polynomial $Q(k)$ associated with the matrix ${\bf W}(k)$, i.e., 
$Q(k;\lambda) = det[\lambda {\bf I} - {\bf W}(k)]$. Therefore, 
$\lambda_{min}(k)$ is a root of the polynomial $Q(k;\lambda)$, i.e., 
solution of the equation
\begin{equation}
Q(k;\lambda) = \sum_{\mu=0}^M C_{\mu}(k) [\lambda(k)]^{\mu} = 0.
\label{eq-poly}
\end{equation}
Hence \cite{koza99,koza00} 
\begin{eqnarray}
V &=& - i \frac{C_{0}'}{C_{1}(0)} \nonumber \\
D &=& \frac{C_{0}'' - 2i C_{1}'(0) V - 2 C_{2}(0) V^{2}}{2 C_{1}(0)} 
\label{eq-Fourier}
\end{eqnarray}
where $C_{\mu}' = [\partial C_{\mu}(k)/\partial k]_{k=0}$. 

For a postulated kinetic scheme, ${\bf W}$ is given. Then the 
expressions (\ref{eq-Fourier}) are adequate for analytical 
derivation of $V$ and $D$ for the given model. However, in order to 
calculate the distributions of the dwell times of the motors, it is 
more convenient to work with the Fourier-Laplace transform, rather 
than the Fourier transform, of the probability densities. Therefore, 
we now derive alternative expressions for $V$ and $D$ in terms of 
the full Fourier-Laplace transform of the probability density. 

Taking Laplace transform of (\ref{eq-FourierP}) with respect to time 
\begin{equation}
\tilde{P}_{\mu}(k,s) = \int_{0}^{\infty} \bar{P}_{\mu}(k,t) e^{-s t}, 
\end{equation}
the matrix form of the master equation reduces to 
$\tilde{{\bf P}}(k,s) = {\bf R}(k,s)^{-1} {\bf P}(0)$ 
where ${\bf R}(k,s) = s{\bf I} - {\bf W}(k)$ 
and ${\bf P}(0)$ is the column vector of initial probabilities. 
Now the characteristic polynomial is the determinant of ${\bf R}(k,s)$ 
which can be expressed as \cite{chemla08}
\begin{equation}
|{\bf R}(k,s)| = \sum_{\mu=0}^M c_{\mu}(k) s^{\mu} = 0.
\label{eq-poly2}
\end{equation}
Equation (\ref{eq-poly2}) is formally similar to (\ref{eq-poly}). 
As expected, we get \cite{chemla08} 
\begin{eqnarray}
V &=& - i \frac{c_{0}'}{c_{1}(0)} \nonumber \\
D &=& \frac{c_{0}'' - 2i c_{1}'(0) V - 2 c_{2}(0) V^{2}}{2 c_{1}(0)} 
\label{eq-FourierLaplace}
\end{eqnarray}

\noindent$\bullet${\bf Example: an M-step unbranched mechano-chemical cycle}

\begin{figure}[ht]
\begin{center}
\includegraphics[angle=-90,width=0.85\columnwidth]{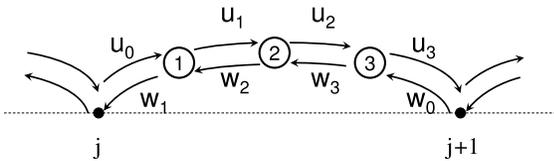}
\end{center}
\caption{An unbranched mechano-chemical cycle of the molecular motor 
with $M = 4$. The horizontal dashed line
shows the lattice which represents the track; $j$ and $j+1$
represent two successive binding sites of the motor. The circles
labelled by integers denote different ``chemical'' states in
between $j$ and $j+1$. (Adapted from fig.7 of ref.\cite{garai09}).
}
\label{fig-fishkolo}
\end{figure}

As an illustrative example, let us consider the unbranched mechano-chemical 
cycle \cite{fisher07} with $M=4$, as shown in fig.\ref{fig-fishkolo}. 
This special value of $M$ is motivated by the typical example of a 
kinesin motor for which the four essential steps in each cycle are as  
follows: 
(i) a substrate-binding step (e.g., binding of an ATP molecule), 
(ii) a chemical reaction step (e.g., hydrolysis of ATP), 
(iii) a product-release step (e.g., release of ADP), and 
(iv) a mechanical step (e.g., power stroke). 

Suppose, The forward transitions take place at rates $u_j$ whereas the 
backward transitions occur with the rates $w_j$.  The average
velocity $V$ of the motor is given by \cite{fisher07}
\begin{equation}
 V=\frac{1}{R_{M}}\biggl[1-\prod^{{M}-1}_{j=0}\biggl(\frac{w_j}{u_j}\biggr)\biggr] 
\label{eq-kolov}
\end{equation}
where
\begin{equation}
R_{M}=\sum^{{M}-1}_{j=0}r_j   
\label{eq-fv3}
\end{equation} 
with
\begin{equation}
r_j=\biggl(\frac{1}{u_j}\biggr)\biggl[1+\sum^{{M}-1}_{k=1}\prod^{k}_{i=1}\biggl(\frac{w_{j+i}}{u_{j+i}}\biggr)\biggr] 
\label{eq-fv4}
\end{equation}
while $D$ is given by
\begin{equation}
 D=\biggl[ \frac{(VS_M+dU_M)}{R^2_M}-\frac{(M+2)V}{2} \biggr]\frac{d}{M}
\end{equation}
where \begin{equation}
 S_M=\sum^{M-1}_{j=0}s_j\sum^{M-1}_{k=0}(k+1)r_{k+j+1}
\end{equation}
and
\begin{equation}
 U_M=\sum^{M-1}_{j=0}u_jr_js_j
\end{equation}
while,
\begin{equation}
 s_j=\frac{1}{u_j}\biggl(1+\sum_{k=1}^{M-1}\prod^{k}_{i=1}\frac{w_{j+1-i}}{u_{j-i}}\biggr).
\end{equation}
For various extensions of this scheme see ref.\cite{fisher07}. 

In the simpler case shown in (\ref{eq-MMlikemot}), 
where $M=2$, and the second step is purely irreversible,  
using the step size ${\ell}$ explicitly (to make the 
dimensions of the expressions explicitly clear), we get 
\begin{eqnarray}
V &=& {\ell} \biggl[\frac{\omega_{1} \omega_{2}}{\omega_{1}+\omega_{-1}+\omega_{2}}\biggr] \nonumber \\
D &=& \frac{\ell^2}{2} \biggl[\frac{(\omega_{1}\omega_{2})-2 (V/\ell)^2}{\omega_{1}+\omega_{-1}+\omega_{2}}\biggr] 
\label{eq-2strir}
\end{eqnarray} 
Note that if, in addition, $\omega_{-1}$ vanishes, i.e., if both the 
steps are fully irreversible, then 
$d/V = \omega_{1}^{-1} + \omega_{2}^{-1}$, 
i.e., the average time taken to move forward by one site is the sum 
of the mean residence time in the two steps of the cycle.

\noindent$\bullet${\bf Effects of branched pathways and inhomogeneities of tracks}

Microtubule-associated proteins and actin-related proteins can give 
rise to inhomogeneities of the tracks for cytoskeletal motors. The 
intrinsic sequence inhomogeneity of DNA and RNA strands can 
nontrivial influences on the motors that use nucleic acid strands 
as their tracks \cite{kafri04}.

\subsection{Jamming on a crowded track: flux-density relation}
In the preceeding subsection we considered lone motor moving 
along a filamentous track. Now we consider a track with heavy 
motor traffic. In principle, unless controlled by some regulatory 
signals, formation of ``traffic jams'' 
\cite{chowdhury00,chowdhury05a,schadschneider10} 
on crowded tracks cannot be ruled out  
\cite{lipowsky06,parmeggiani04,popkov03,evans03}. 
These phenomena are formally similar to systems of sterically 
interacting self-propelled particles. One of the simplest theoretical 
models for such systems is the totally asymmetric simple exclusion 
process (TASEP). In fact, TASEP was originally inspired by traffic 
of ribosomes on a mRNA track during translation 
\cite{macdonald68,macdonald69,chou11,haar12}.  
However, in recent years it has found application also in the 
context of traffic-like collective movements of ribosomes in more 
realistic models \cite{basu07,sharma11b} as well as many other types 
of molecular motors, e.g., kinesins \cite{nishinari05,greulich07}, 
RNA polymerase \cite{tripathi08,klumpp11}, growth of fungal hyphae 
\cite{sugden07}, growth of bacterial flagella \cite{schmitt11}, 
crowding of depolymerizing kinesins at the tip of a microtube 
\cite{reese11}, etc.

\subsection{\bf Information processing machines: fidelity versus power and efficiency}

Power and efficiency are most appropriate quantities to characterize 
the performance 
of porters, rowers, sliders, etc. But, not all machines fall in these 
categories. For information processing machines, which we consider in 
this subsection, fidelity of information transfer is more important 
than power and efficiency. 

The sequence of the monomeric subunits of a polynucleotide or that of a 
polypeptide is dictated by that of the corresponding template strand. 
During the polymerization process, after preliminary tentative selection, 
the selected monomeric subunit is subjected to several levels of checks 
by the quality control system of the polymerizing machinery. Only after 
a selected subunit is screened by such a stringent test, it is covalently 
bonded to the elongating biopolymer. 

Discrimination of monomers based merely on the differences of free-energy 
gains cannot account for the observed high fidelity of these processes. 
For example, ``kinetic proofreading'' \cite{hopfield74,ninio75} 
has been invoked to explain the kinetic processes that contribute to 
the high level of translational fidelity. 
Following three features are {\it essential} for kinetic 
proofreading \cite{yarus92a}:\\
(i) formation of an initial enzyme-reactant complex, \\
(ii) a strongly forward driven step that results in a high-energy
intermediate complex, and\\
(iii) one or more branched pathways along which dissociation of
the enzyme-reactant complex, and rejection of the reactant, is
possible before the complex gets an opportunity to make the final
transition to yield the product.

Let is consider the kinetic scheme shown below 
\begin{eqnarray}
M_{j}+P_{n}+S \rightleftharpoons I_{1} \rightarrow \mathop{I_{2}}_{\downarrow} \rightarrow M_{j+1}+P_{n+1} \nonumber \\
\label{eq-MMlikeproof}
\end{eqnarray}
where $M_j$ denotes the polymerase motor located at the discrete 
position $j$ on its template, $P_{n}$ represents the elongating 
biopolymer consisting of $n$ subunits and $S$ is a single subunit.
Note that the scheme shown in (\ref{eq-MMlikeproof}) is, at least, 
formally an extension of (\ref{eq-MMlikemot}) where an extra  
intermediate state $I_2$ and a branched path from $I_2$ have been 
added. This scheme is one of the simplest possible implementations 
of kinetic proofreading.

Kinetic proofreading leads to futile cycles in which at least part of 
the input free energy is dissipated without elongating the nascent 
polypeptide by an amino acid monomer. The effects of the consequent  
loose mechano-chemical coupling \cite{oosawa00} of the translational 
machinery on the rate of polypeptide synthesis has been investigated 
\cite{sharma10,sharma11a,sharma11b}. 

Occasionally wrong monomers escape detection in spite of elaborate 
selection procedure. In case of DNA replication, the polymerase normally 
detects the error immediately and corrects it before elongating it further. 
For this purpose, the nascent DNA is transferred 
to another specific site on the same polymerase where the wrong monomer is 
cleaved before transferring the nascent DNA back to the site of elongation 
activity. The interplay of the two contradictory activities of elongation 
and shortening of a nascent DNA exposes leads to the coupling of their 
corresponding rates in the overall rate of DNA polymerization by a DdDP  
\cite{sharma12}.

\subsection{\bf Beyond average: dwell time distribution (DTD)}

Two motors with identical average velocities may exhibits widely 
different types of fluctuations. 
Suppose the successive mechanical steps are taken by the motor at 
times $t_{1}, t_{2},...,t_{n-1}, t_{n}, t_{n+1},...$. Then, the 
time of dwell before the $k$-th step is defined by 
$\tau_{k} = t_{k}-t_{k-1}$. In between successive steps, the motor 
may visit several ``chemical'' states and each state may be visited 
more than once. But, the purely chemical transitions would not be 
visible in a mechanical experimental set up that records only its 
position. The number of visits to a given state and the duration of 
stay in a state in a given visit are random quantities.

In order to appreciate the origin of the fluctions in the dwell times, 
let us consider the simple $N$-step kinetics: 
\begin{equation}
M_{1} \rightleftharpoons M_{2} \rightleftharpoons M_{3} ...\rightleftharpoons M_{j} \rightleftharpoons ...\rightleftharpoons M_{N}
\label{eq-cycle}
\end{equation}
Suppose $t_{\mu,\nu}$ is the duration of stay of the motor in the 
$\mu$-th state during its $\nu$-th visit to this state. If $\tau$ 
is the dwell time, then 
\begin{equation}
\tau = \sum_{\mu=1}^{N} \sum_{\nu=1}^{n_{\mu}} t_{\mu,\nu}
\end{equation}
where $n_{\mu}$ is the number of visits to the $\mu$-th state.
It is straightforward to check that 
\begin{eqnarray} 
<\tau> &=& \sum_{\mu=1}^{N} <n_{\mu}><t_{\mu}> \nonumber \\
\end{eqnarray}
where $<n_{\mu}>$ is the avarege number of visitits to the $\mu$-th 
state and $<t_{\mu}>$ is the avarege time of dwell in the $\mu$-th 
state is a single visit to it. More interestingly \cite{moffitt10}, 
\begin{eqnarray} 
&&<\tau^2>-<\tau>^2 = \sum_{\mu=1}^{N} (<t_{\mu}^2> - <t_{\mu}>^2) <n_{\mu}> \nonumber \\ 
&+& \sum_{\mu=1}^{N} (<n_{\mu}^2> - <n_{\mu}>^2) <t_{\mu}>^2 \nonumber \\
&+& 2 \sum_{\mu > \nu} (<n_{\mu} n_{\nu}>-<n_{\mu}><n_{\nu}>)<t_{\mu}><t_{\nu}> \nonumber \\
\label{eq-correl}
\end{eqnarray}
The first and second terms on the right hand side of (\ref{eq-correl}) 
capture, respectively, the fluctuations in the lifetimes of the 
individual states and that in the number of visits to a kinetic state. 
Note that the number of visits to a particular state depends on the 
number of visits to the neighboring states on the kinetic diagram; 
this interstate correlation is captured by the third term on the 
right hand side of (\ref{eq-correl}).

Several different analytical and numerical techniques have been 
developed for calculation of the dwell time distribution 
\cite{shaevitz05,liao07,linden07,chemla08}. 
Since the dwell time 
is essentially a first passage time \cite{rednerbook}, an absorbing 
boundary method \cite{liao07} has been used.

\subsubsection{DTD for a motor that never steps backward}

As an example, we consider again the simple scheme (\ref{eq-MMlikemot}). 
In this case, the DTD is 
\begin{equation}
f(t) = \biggl( \frac{\omega_{1}\omega_{2}}{\omega_{-}-\omega_{+}}\biggr)(e^{-\omega_{+}t} - e^{-\omega_{-}t}) 
\label{eq-coupledft}
\end{equation}
where 
\begin{equation} 
\omega_{\pm} = \frac{\omega_{1}+\omega_{-1}+\omega_{2}}{2}\pm\biggl[\sqrt{\frac{(\omega_{1}+\omega_{-1}+\omega_{2})^2}{4}-\omega_{1}\omega_{2}}\biggr]
\label{eq-omegapm}
\end{equation}
In the special case $\omega_{-1} = 0$, $\omega_{+} = \omega_{1}$ and 
$\omega_{-} = \omega_{2}$
and, hence, 
\begin{equation}
f(t) = \biggl( \frac{\omega_{1}\omega_{2}}{\omega_{2}-\omega_{1}}\biggr)(e^{-\omega_{1}t} - e^{-\omega_{2}t}) 
\label{eq-uncoupledft}
\end{equation} 
Similar sum of exponentials for DTD have been derived also for machines  
with more complex mechano-chemical kinetics (see, for example, 
refs.\cite{garai09,tripathi09,sharma11a}).

It is possible to establish on general grounds that, for a motor 
with $N$ mechano-chemical kinetic states like (\ref{eq-cycle}), 
the DTD is a sum of $n$ exponentials of the form \cite{moffitt10}
\begin{equation}
f(t) = \sum_{j=1}^{N} C_{j} e^{-\omega_{j}t}
\label{eq-sumexpo1}
\end{equation}
where $N-1$  of the $N$ coefficients $C_{j}$ ($1 \leq j \leq N$) 
are independent of each other because of the constraint imposed by 
the normalization condition for the distribution $f(t)$. Also note 
that the prefactors $C_{j}$ can be both positive or negative while 
$\omega_{j} > 0$ for all $j$. 

If we consider an even more special case of the scheme (\ref{eq-MMlikemot}) 
where $\omega_{-1}=0, \omega_{1}=\omega_{2}=\omega$, i.e., all the 
steps are completely irreversible and take place at the same rate 
$\omega$, the DTD becomes the Gamma-distribution 
\begin{equation}
f(t) = \frac{\omega^{N} t^{N-1} e^{-\omega t}}{\Gamma(N)} 
\end{equation}
where $\Gamma(N)$ is the gamma function with $N=2$.
Interestingly, for the Gamma-distribution, the randomness parameter 
\cite{schnitzer95} (also called the Fano factor \cite{fano47})  
\begin{equation} 
r = (<\tau^2>-<\tau>^2)^{1/2}/<\tau> 
\label{eq-fano} 
\end{equation} 
is exactly given by $r=1/N$. Can the experimentally measured DTD 
be used to determine the number of states $N$? Unfortunately, 
for real motors, (i) not each step of a cycle is fully irreversible, 
(ii) the rate constants for different steps are not necessarily identical, 
(iii) branched pathways are quite common. Consequently, $1/r$ may 
provide just a bound on the rough estimate of $N$. 

Can one use the general form (\ref{eq-sumexpo1}) of DTD to extract 
all the rate constants for the kinetic model by fitting it with the 
experimentally measured DTD? The answer is: NO.
First, even if a good estimate of $N$ is available, the number of 
parameters that can be extracted by fitting the experimental DTD 
data to (\ref{eq-sumexpo1}) is $2N-1$ ($n$ values of $\omega_{j}$ 
and $N-1$ values of $C_{j}$). 
On the other hand, the number of possible rate constants may be much 
higher \cite{moffitt10}. For example, if transitions from every kinetic 
state to every other kinetic state is allowed, the total number of rate 
constants would be $N(N-1)$. In other words, in general, the kinetic 
rate constants are underspecified by the DTD. 
Second, as the expression (\ref{eq-coupledft}) for the DTD of the 
example (\ref{eq-MMlikemot}) shows expplicitly, the $\omega$'s that appear 
in the exponentials may be combinations of the rate constants for the 
distinct transitions in the kinetic model. It is practically impossible 
to extract the individual rate constants from the estimated $\omega$'s  
unless any explicit relation like (\ref{eq-omegapm}) between the estimated 
$\omega$'s and actual rate constants is {\it apriori} available.

\subsubsection{Conditional DTD for motors with both forward and backward stepping}

For motors which can step both forward and backward, more relevant  
information on the kinetics of a motor are contained in the 
{\it conditional} dwell time distribution \cite{tsygankov07}. 
If $n_{+}$ and $n_{-}$ are the numbers of forward and backward steps, 
respectively, then for large $n = n_{+} + n_{-}$, the corresponding 
{\it step splitting} probabilities are $\Pi_{+} = n_{+}/n$ and 
$\Pi_{-} = n_{-}/n$. The dwell times {\it before} a forward step and 
before a backward step can be measured separately. Hence, the 
{\it prior dwell times} $\tau_{+}$ and $\tau_{-}$ can be obtained 
by restricting the summations in 
\begin{equation}
\tau_{\pm}^{\leftarrow} = \frac{1}{n_{\pm}} \sum^{\pm} \tau_{k}
\end{equation}
to just forward (+) or just backward (-) steps, respectively. In 
terms of splitting probabilities and prior dwell times, the 
mean dwell time $\langle \tau \rangle$ can be expressed as 
\begin{equation}
\langle \tau \rangle = \Pi_{+} \tau_{+}^{\leftarrow} + \Pi_{-} \tau_{-}^{\leftarrow}
\end{equation}

Compared to the prior dwell times, more detailed information on 
the stepping statistics is contained in the four {\it conditional 
dwell times}, which are defined as follows: 
\begin{eqnarray} 
\tau_{++} &=& {\rm dwell ~time ~between ~a ~+ ~step ~followed ~by ~a ~+ ~step} \nonumber \\
\tau_{+-} &=& {\rm dwell ~time ~between ~a ~+ ~step ~followed ~by ~a ~- ~step} \nonumber \\
\tau_{-+} &=& {\rm dwell ~time ~between ~a ~- ~step ~followed ~by ~a ~+ ~step} \nonumber \\
\tau_{--} &=& {\rm dwell ~time ~between ~a ~- ~step ~followed ~by ~a ~- ~step} \nonumber \\
\end{eqnarray} 
It is helpful to introduce {\it pairwise step probabilities} 
$\Pi_{++}, \Pi_{+-}, \Pi_{-+}, \Pi_{--}$. Note that 
$\Pi_{++}+\Pi_{+-} = 1$, and $\Pi_{-+} + \Pi_{--} = 1$. 
Neglecting finite time corrections, 
\begin{eqnarray} 
\Pi_{++} = n_{++}/(n_{++}+n_{+-}),~~\Pi_{+-} = n_{+-}/(n_{++}+n_{+-})\nonumber \\
\Pi_{-+} = n_{-+}/(n_{-+}+n_{--}),~~\Pi_{--} = n_{--}/(n_{-+}+n_{--})\nonumber \\
\end{eqnarray} 
Hence 
\begin{eqnarray}
\tau_{+}^{\leftarrow} &=& \Pi_{++} \tau_{++} + \Pi_{+-} \tau_{-+} \nonumber \\
\tau_{-}^{\leftarrow} &=& \Pi_{-+} \tau_{+-} + \Pi_{--} \tau_{--} 
\end{eqnarray} 
Defining the {\it post dwell times} $\tau_{\pm}^{\rightarrow}$ in 
a fashion similar to that used for defining the {\it prior dwell 
times}, we get 
\begin{eqnarray}
\tau_{+}^{\rightarrow} &=& \Pi_{++} \tau_{++} + \Pi_{+-} \tau_{+-} \nonumber \\
\tau_{-}^{\rightarrow} &=& \Pi_{-+} \tau_{-+} + \Pi_{--} \tau_{--} 
\end{eqnarray} 
and 
\begin{equation}
\langle \tau \rangle = \Pi_{+} \tau_{+}^{\rightarrow} + \Pi_{-} \tau_{-}^{\rightarrow}
\end{equation}
This formalism has been applied, for example, single-headedkinesin 
KIF1A \cite{garai11}.

\section{A summary of experimental techniques: ensemble versus single-machine} 

The experimental techniques for probing the structure and dynamics of
molecular machines can be divided broadly into two groups: 
(i) ensemble-averaged, (ii) single-machine. 

X-ray diffraction and electron microscopy, two of the most powerful 
biophysical techniques, yield the structure of molecular machines 
averaged over an ensemble. Electron microscopy is a powerful alternative 
for the determination of structures of those macromolecules whose 
crystals are not available. Cryo-electron microscopy 
\cite{glaeser08,frank11} combines the technique of electron microscopy 
with cryogenics-based sample preparation.  

The single-molecule techniques
\cite{zlatanova06,ritort06,cornish07,kapanidis09,moffitt08,greenleaf07}
can be further classified into two groups: (i) methods of imaging, and
(ii) methods of manipulation. These are not mutually exclusive and can 
be probed by a single experimental set up.
The most direct single-molecule imaging is carried out by fluorescence-based 
optical microscopy. Traditional diffraction-limited optical microscopy 
provides a ``hazy'' image of the machine. However, modern optical nanoscopy 
has broken the resolution limit imposed by diffraction \cite{schermelleh10}.
Fluorescence microscopy provides a glimpse (howsoever hazy) of single
molecules. Imaging a fluorescently labelled molecular machine in real
time enables us to study its dynamics just as ecologists use ``radio
collars'' to track individual animals. 

The techniques developed for the mechanical manipulation of a single 
machine can be classified on the basis of tranducers used; the table 
below provides a summary. 

\centerline{\framebox{Mechanical manipulation of a single biomolecule}}

\centerline{$\swarrow$ ~~ $\searrow$}
\centerline{{\framebox{Mechanical transducers}} ~~~~~~~ {\framebox{Field-based t
ransducers}}}
\centerline{$\swarrow$ ~~ $\searrow$ ~~~~~~~~~~~~~~~~~~~~~~~~~~ $\swarrow$ ~~ $\searrow$ }

\centerline{~~~~~{\framebox{SFM}}~ {\framebox{Micro-needle}}~~~~~{\framebox{EM-field}}~ {\framebox{Flow-field}}   }

\centerline{~~~~~~~~~~~~~~~~~~~~~~~ $\swarrow$ ~~ $\searrow$ }

\centerline{~~~~~~~~~~~~~~~~~~~~~~~~~{\framebox{ Electric field}}~ {\framebox{Ma
gnetic field}}   }


\section{Solving inverse problem by probabilistic reverse engineering: from data to model} 

A discrete kinetic model of a molecular motor can be regarded as a 
network where each node represents a distinct mechano-chemical state. 
The directed links denote the allowed transitions. Therefore, such a 
model is unambiguously specified in terms of the following parameters: 
(i) the total {\it number} $N$ of the nodes, 
(ii) the $N \times N$ matrix whose elements are the {\it rates} of 
the transitions among these states; a vanishing rate indicates a 
forbidden direct transition. 

In the preceeding sections we handled the ``forward problem'' by starting 
with a model that is formulated on the basis of {\it apriori} hypotheses 
which are, essentially, educated guess as to the mechano-chemical kinetics 
of the motor. Standard theoretical treatments of the model yields data on 
various aspects on the modelled motor; this approach is expressed below 
schematically.
\centerline{{\framebox{Theoretical model}} ~~$\rightarrow$~~ {\framebox{Experimental data}} }
Consistency between theoretical prediction and expermental data validates  
the model. However, any inconsistency between the two indicates a need to 
modify the model.  

In most real situations the numerical values of the rate constants of 
the kinetic model are not known apriori. In principle, these can be 
extracted by analyzing the experimental data in the light of the model. 
However, not only the rate constants but also the number of states and 
the overall architecture of the mechano-chemical network as well as 
the kinetic scheme postulated by the model may be uncertain. In that 
case, the experimental data should be utilized to ``select'' the most 
appropriate model from among the plausible ones. In fact, more than one 
model, based on different hypotheses, may appear to be consistent with 
the same set of experimental data within a level of accuracy. The 
experimental data can be exploited at least to ``rank'' the models in 
the order of their success in accounting for the same data set. 

The ``inverse problem'' of inferring the model from the observed 
experimental data has to be based on the theory of probability. 
Such ``statistical inference'' \cite{balding11} can be drawn by 
following methods developed by statisticians over the last one century. 
Inferring the complete network of mechano-chemical states and kinetic 
scheme of a molecular machine from its observed properties is reminiscent 
of inferring the operational mechanism of a given functioning macroscopic 
machine by ``reverse engineering''. This inverse problem, which is 
the main aim of this section, is expressed below schematically. 
\centerline{{\framebox{Theoretical model}} ~~$\leftarrow$~~ {\framebox{Experimental data}} }

It is worth emphasizing that both the directions of investigations,
i.e. the forward problem and the inverse problem are equally important
and complementary to each other \cite{kell03}.

\subsection{Frequentist versus Bayesian approach}

Suppose, $\vec{m}$ be a column vector whose $M$ components are the $M$
parameters of the model, i.e., the transpose of $\vec{m}$ is
$\vec{m}^{T} = (m_1,m_2,...,m_M)$
Let the data obtained in $N$ observations of this model are represented
by the $N$-component column vector $\vec{d}$ whose transpose is
$\vec{d}^{T} = (d_1,d_2,...,d_N)$. Our ``inverse problem'' is to infer 
information on $\vec{m}$ from the observed $\vec{d}$.
The philosophy underlying the frequentist approach, i.e., approaches
based on maximum-likelihood (ML) estimation and the Bayesian approach 
for extracting these information are different in spirit, as we explain 
in the next two subsubsections \cite{cowan07}.

For simplicity, let us assume that a device has only two possible
distinct states denoted by ${\cal E}_{1}$ and ${\cal E}_{2}$.
\begin{equation}
{\cal E}_{1} \mathop{\rightleftharpoons}^{k_f}_{k_r} {\cal E}_{2}
\label{eq-stateseq}
\end{equation}
Let us imagine that we are given the actual sequence of the states,
over the time interval $0 \leq t \leq T$, generated by the Markovian
kinetics of the device. But, the magnitudes of the rate constants
$k_f$ and $k_r$ are not supplied. We'll now formulate a method,
based on ML analysis \cite{ andrec03} to extract the numerical
values of the parameters $k_f$ and $k_r$.

Suppose $t_{j}^{(1)}$ and $t_{j}^{(2)}$ denote the time interval of
the $j$-th residence of the device in states ${\cal E}_{1}$ and
${\cal E}_{2}$, respectively. Moreover, suppose that the device
makes $N_1$ and $N_2$ visits to the states ${\cal E}_{1}$ and
${\cal E}_{2}$, respectively, and $N = N_1+N_2$ is the total
number of states in the sequence. Therefore, total time of dwell
in the two states are
$T_{1} = \sum_{j=1}^{N_1} t_{j}^{(1)}$ and
$T_{2} = \sum_{j=1}^{N_2} t_{j}^{(2)}$
where $T_{1}+T_{2} = T$.

Since the dwell times are exponentially distributed for a Poisson
process, the likelihood of any state trajectory $S$ is the 
conditional probability 
\begin{eqnarray}
P(S|\underline{k_f,k_r}) &=& \biggl(\Pi_{j=1}^{N_1} k_f e^{-k_f t_j^{(1)}}\biggr) \nonumber \\
&&\biggl(\Pi_{j=1}^{N_2} k_r e^{-k_r t_j^{(2)}}\biggr) \nonumber \\
&=& \biggl(k_f^{N_1} e^{-k_f T_1}\biggr)\biggl(k_r^{N_2} e^{-k_r T_2}\biggr) 
\label{eq-likeli}
\end{eqnarray}

\subsubsection{Maximum-likelihood estimate}

ML approcah \cite{myung03} is based on finding the estimates
of the set of model parameters that corresponds to the maximum of
the likelihood $P(\vec{d}|\underline{\vec{m}})$ for a fixed set of
data $\vec{d}$.

For the kinetic scheme shown in equation (\ref{eq-stateseq}), the
the ML estimates of $k_f$ and $k_r$ are obtained by using (\ref{eq-likeli}) 
in
$d[ln P(S|\underline{k_f,k_r})]/dk_f=0 = d[ln P(S|\underline{k_f,k_r})]/dk_r$.
It is straightforward to see \cite{andrec03} that these estimates are
$k_f=N_{1}/T_{1}$ and $k_r=N_{2}/T_{2}$.

\subsubsection{Bayesian estimate}

For drawing statistical inference regarding a kinetic model, the
Bayesian approach has gained increasing popularity in recent
years \cite{raeside72a,raeside72b,ulrych01,scales01,eddy04a,kou05b}.
The areas of research where this has been applied successfully 
include various biological processes in, for example, genetics 
\cite{schoemaker99,beaumont04}, biochemistry \cite{golightly11},
cognitive sciences \cite{kruschke10}, ecology \cite{ellison04}, etc. 

In the Bayesian method there is no logical distinction between the
model parameters and the experimental data; in fact, both are
regarded as random. The only distinction between these two types
of random variables is that the data are observed variables whereas
the model parameters are unobserved. The problem is to estimate the 
{\it probability distribution} of the model parameters from the 
distributions of the observed data.

The Bayes theorem states that
\begin{equation}
P(\vec{m}|\underline{\vec{d}}) = \frac{P(\vec{d}|\underline{\vec{m}}) P(\vec{m})}{P(\vec{d})}
\label{eq-bayes1}
\end{equation}
where $P(\vec{d})$ can be expressed as
\begin{equation}
P(\vec{d}) = \int P(\vec{d}|\underline{\vec{m}}) P(\vec{m}) d\vec{m}
\label{eq-bayes2}
\end{equation}
The {\it likelihood} $P(\vec{d}|\underline{\vec{m}})$ is the conditional
probability for the observed data, given a set of particular values of
the model parameters, that is predicted by the kinetic model.  However, 
implementation of this scheme also requires $P(\vec{m})$ as input.
In Bayesian terminology $P(\vec{m})$ is called the {\it prior} because
this probability is {\it assumed} {\it apriori} by the analyzer
{\it before} the outcomes of the experiments have been analyzed.
In contrast, the left hand side of equation (\ref{eq-bayes1}) gives the
{\it posterior} probability, i.e., after analyzing the data.

Thus, an experimenter learns from the Bayesian analysis of the data.
Such a learning begins with an input in the form of a prior probability;
the choice of the prior can be based on physical intuition, or general
arguments based, for example, on symmetries. Prior choice can become
simple if some experience have been gained from previous measurements.
Often an uniform distribution of the model parameter(s) is assumed over 
its allowed range if no additional information is available to bias its
choice. To summarize, Bayesian analysis needs not just the likelihood
$P(\vec{d}|\underline{\vec{m}})$ but also the prior $P(\vec{m})$.

For the kinetic scheme shown in equation (\ref{eq-stateseq}), the Bayes'
theorem (\ref{eq-bayes1}) takes the form
\begin{eqnarray}
P(k_f,k_r|\underline{S}) &=& \frac{P(S|\underline{k_f,k_r}) P(k_f,k_r)}{P(S)} \nonumber \\
&=& \frac{P(S|\underline{k_f,k_r}) P(k_f,k_r)}{\sum_{k_f',k_r'} P(S|\underline{k_f',k_r'}) P(k_f',k_r')}
\label{eq-bayes3}
\end{eqnarray}
We now assume a uniform prior, i.e., constant for positive $k_f$ and $k_r$,
but zero otherwise. Then, $P(k_f,k_r|\underline{S})$ is proportional to
the likelihood function $P(S|\underline{k_f,k_r})$ (within a normalization
factor). Normalizing, we get
\begin{equation}
P(k_f,k_r|\underline{S}) =
\biggl[\frac{T_{1}^{N_1+1}}{\Gamma(N_1+1)}k_f^{N_1} e^{-k_f T_1}\biggr]\biggl[\frac{T_{2}^{N_2+1}}{\Gamma(N_2+1)}k_r^{N_2} e^{-k_r T_2}\biggr]
\label{eq-bayes4}
\end{equation}
The mean of $k_f$ is $N_1+1)/T_1$ whereas the most-likely estimate is
$N_1/T_1$. Similarly, the mean and most probable estimates of $k_r$
are obtained by replacing the subscripts $1$ by subscripts $2$.
Moreover, the variance of $k_f$ and $k_r$ are $(N_1+1)/T_1^2$ and
$(N_2+1)/T_2^2$, respectively.

\subsubsection{Hidden Markov Models}

The actual sequence of states of the motor, generated by the underlying
Markovian kinetics, is not directly visible. For example, a sequence of
states that differ ``chemically'' but not mechanically do not appear as
distinct on the recording of the position of the motor in a single motor 
experiment.
This problem is similar to an older problem in cell biology: ion-channel
kinetics \cite{ball92,hawkes04}. 
Current passes through the channel only when it is in the ``open'' state.
However, if the channel has more than one distinct closed states, the
recordings of the current reveals neither the actual closed state in
which the channel was nor the transitions between those closed states when
no current was recorded.

Hidden Markov Model (HMM) \cite{eddy04b,rabiner89,talaga07,vogl10}
has been applied to analyze FRET trajectories \cite{mckinney06,lee09}, 
stepping recordings of molecular motors \cite{mullner10,syed10}, and 
actomyosin contractile system \cite{smith01}
to extract kinetic information.

For a pedagogical presentation of the main ideas behind HMM, we start
with the kinetic scheme shown in (\ref{eq-bayes1}) and
a simple, albeit unrealistic, situation and then by gradually
adding more and more realistic features, explain the main concepts in
a transparent manner \cite{andrec03}. First, suppose that the actual 
sequence of states (trajectory) itself is visible; this case can be 
analyzed either by the ML-analysis of by Bayesian approach both of 
which we have presented above. We now relax the strong assumption about 
the trajectory and proceed as below.

\noindent$\bullet${\bf If state trajectory is hidden and visible trajectory is noise-free}

The sequence of states of the device is, as before, generated by a 
Markov processs which is {\it hidden}. Suppose the device emits 
photons from time to time that are detected by appropriate 
photo-detectors. For simplicity, we assume just two detection 
channels labelled by $1$ and $2$. For the time being, we also 
assume a perfect one-to-one correspondence between the state of the 
light emitting device and the channel that detects the photon.
If the channel $1$ ($2$) clicks then the light emitting device was
in the state ${\cal E}_{1}$ (${\cal E}_{2}$) at the time of emission.
The interval $\Delta t_{j} = t_{j+1}-t_{j}$ between the arrival of
the $j$-th and $j+1$-th photons ($1 \leq j \leq N$) is random.

Thus, from the photo-detectors we get a visible sequence of the 
channel index (a sequence made of a binary alphabet) which we call 
``noiseless photon trajectory'' \cite{andrec03}. 
The sequence of states in the noiseless photo trajectory is also 
another Markov chain that is conventionally referred to as  the 
``random telegraph process''. Note that the photon detected by 
channel 1 (or, channel 2) can take place at any instant during the 
dwell of the device in state 1 (or, state 2). Therefore, the sequence 
of states in the noiseless photon trajectory does not reveal the 
actual instants of transition from one state of the device to another.

Since the noiseless phton trajectory corresponds to a random telegraph 
process, the transition probabilities for this process are 
\begin{eqnarray}
P({\cal E}_{1}|\underline{{\cal E}_{1};k_f,k_r,\Delta t_{j}}) &=& \frac{k_r}{k_f+k_r} + \frac{k_f}{k_f+k_r} e^{-(k_f+k_r)\Delta t_j} \nonumber \\
P({\cal E}_{1}|\underline{{\cal E}_{2};k_f,k_r,\Delta t_{j}}) &=& \frac{k_r}{k_f+k_r} [1 - e^{-(k_f+k_r)\Delta t_j}] \nonumber \\
P({\cal E}_{2}|\underline{{\cal E}_{1};k_f,k_r,\Delta t_{j}}) &=& \frac{k_f}{k_f+k_r} [1 - e^{-(k_f+k_r)\Delta t_j}] \nonumber \\
P({\cal E}_{2}|\underline{{\cal E}_{2};k_f,k_r,\Delta t_{j}}) &=& \frac{k_f}{k_f+k_r} + \frac{k_r}{k_f+k_r} e^{-(k_f+k_r)\Delta t_j} \nonumber \\
\label{eq-condprob}
\end{eqnarray}
where $P({\cal E}_{\mu}|\underline{{\cal E}_{\nu};k_f,k_r,\Delta t_{j}})$
is the conditional probability that state of the device is ${\cal E}_{\mu}$
given that it was in the state ${\cal E}_{\nu}$ at a time $\Delta t$ earlier.

The likelihood of the visible data sequence $\{V\}$ is now given by
\begin{equation}
P(\{V\}|k_f,k_r) = P(V_1|k_f,k_r) \Pi_{j=1}^{N-1} P(V_{j+1}|V_{j};k_f,k_r,\Delta t_j)
\label{eq-nonoise}
\end{equation}
where the first factor on the right hand side is the initial probability
(usually assumed to be the equilibrium probability). The transition
probabilities on the right hand side of equation (\ref{eq-nonoise}) are
the conditional probabilities given in equation (\ref{eq-condprob}).
Unlike the previous simpler case, where the state sequence itself was
visible, no analytical closed-form solution is possible in this case.
Nevertheless, analysis can be carried out numerically.

\noindent$\bullet${\bf If state trajectory is hidden and visible trajectory is noisy}

In the preceeding case of a noise-free photon trajectory, we assumed that
from the channel index we could get perfect knowledge of the state of the
emitting device. More precisely, the conditional probabilities were
\begin{eqnarray}
P(1|\underline{{\cal E}_{1}}) &=& 1\nonumber \\
P(1|\underline{{\cal E}_{2}}) &=& 0 \nonumber \\
P(2|\underline{{\cal E}_{1}}) &=& 0 \nonumber \\
P(2|\underline{{\cal E}_{2}}) &=& 1 \nonumber \\
\label{eq-emissionprob1}
\end{eqnarray}

However, in reality, background noise is unavoidable. Therefore, if a 
photon is detected by the channel $1$, it could have been emitted by 
the device in its state ${\cal E}_{1}$ (i.e., it is, indeed, a signal 
photon) or it could have come from the background (i.e., it is a noise 
photon). Suppose $p_{s}$ is the probability that the detected photon 
is really a signal that has come from the emitting device. Suppose 
$p_{b1}$ is the probability of arrival of a background photon in the 
channel 1. The probability that a background photon arrives in channel 
2 is $1-p_{b1}$. Then \cite{andrec03},
\begin{eqnarray}
E(1|\underline{{\cal E}_{1}}) &=& p_s + (1-p_s) p_{b1} \nonumber \\
E(2|\underline{{\cal E}_{1}}) &=& 1- P(1|\underline{{\cal E}_{1}}) = (1-p_s) (1-p_{b1}) \nonumber \\
E(1|\underline{{\cal E}_{2}}) &=& (1-p_s) p_{b1} \nonumber \\
E(2|\underline{{\cal E}_{2}}) &=& 1 - P(1|\underline{{\cal E}_{2}}) = p_s + (1-p_s) (1-p_{b}) \nonumber \\
\label{eq-emissionprob2}
\end{eqnarray}
Thus, in this case, the relation between the states of the hidden and 
visible states is not one-to-one, but one-to-many.Therefore, given a 
hidden state of the device, a set of ``emission probabilities'' 
determine the probability of each possible observable state; these 
are listed in equations (\ref{eq-emissionprob2}) for the device 
(\ref{eq-stateseq}).

\noindent$\bullet${\bf HMM: formulation for a generic model of molecular motor}

On the basis of the simple example of a 2-state system presented above, 
we conclude that, for data analysis based on a HMM four key ingredients 
have to be specified:\\
(i) the alphabet of the ``visible'' sequence $\{\mu\}$ ($1 \leq \mu \leq N$),
i.e., $N$ possible distinct visible states;
(ii) the alphabet of the ``hidden'' Markov sequence $\{j\}$
($1\leq j \leq M$), i.e., $M$ allowed distinct hidden states,
(iii) the hidden-to-hidden {\it transition} probabilities $W(j\to k)$, and
(iv) hidden-to-visible {\it emission} probabilities $E(j \to \mu)$.
In addition to the transition probabilities and emission probabilities,
which are the parameters of the model, the HMM also needs the initial
state of the hidden variable(s) as input parameters.

\centerline{Visible: {\framebox{V$_0$}}~~~~~~~~~{\framebox{V$_1$}}~~~~~...~~~~~{\framebox{V$_t$}}~~~~~~~~~ {\framebox{V$_T$}} }
~~~~~~~~~~~~~~$\Uparrow$~~~~~~~~~~~~~$\Uparrow$~~~~~~~~~~~~~~~~$\Uparrow$~~~~~~~~~~~~~$\Uparrow$\\
\centerline{Hidden: {\framebox{H$_0$}}~~~$\rightarrow$~~~~{\framebox{H$_1$}}~$\rightarrow$~...$\rightarrow$~{\framebox{H$_t$}}~~~$\rightarrow$~~~~{\framebox{H$_T$}} }

Suppose $P(\{V\}|\underline{HMM,\{\lambda\}})$ denotes the probability that an
HMM with parameters $\{\lambda\}$ generates a visible sequence $\{V\}$.
Then,
\begin{equation}
P(\{V\}|\underline{HMM,\{\lambda\}}) = \sum_{\{H\}} P(\{V\}|\underline{\{H\};\{\lambda\}}) P(\{H\}|\underline{\{\lambda}\})
\label{eq-HMM1}
\end{equation}
where $P(\{H\}|\underline{\{\lambda\}})$ is the conditional probability 
that the HMM generates a hidden sequence $\{H\}$ for the given parameters 
$\{\lambda\}$ and $P(\{V\}|\underline{\{H\};\{\lambda\}})$ is the 
conditional probability that, given the hidden sequence $\{H\}$ (for 
parameters $\{\lambda\}$) the visible sequence $\{V\}$ would be obtained.

Once $P(\{V\}|\underline{HMM,\{\lambda\}})$ is computed, the parameter set
$\{\lambda\}$ are varied to maximize
$P(\{V\}|\underline{HMM,\{\lambda\}})$ (for the convenience of numerical
computation, often ln$P(\{V\}|\underline{HMM,\{\lambda\}})$ is maximized.
The total number of possible hidden sequences of length $T$ is $T^{MN}$.
In order to carry out the summation in equation (\ref{eq-HMM1}) one has 
to enumerate all possible hidden sequences and the corresponding 
probabilities of occurrences. A successful implementation of the HMM 
requires use of an efficienct numerical algorithm; the {\it Viterbi 
algorithm} is one such candidate.

In case of a molecular motor, the ``chemical states'' are not visible
in a single molecule experiment. Moreover, even its mechanical state
that is ``visible'' in the recordings may not be its true position
because of (a) measurement noise, and (b) steps missed by the detector.
Let is denote the ``visible'' sequence by the recorded positions
$\{Y\}$ whereas the hidden sequence is the composite mechano-chemical
states $\{X,C\}$ where $X$ and $C$ denote the true position and chemical
state, respectively. The transition probabilities are denoted by
$W(X_{t-1},C_{t-1} \to X_t,C_t)$
One possible choice for the emission probabilities $E$ is \cite{mullner10}
\begin{equation}
E(X_t \to Y_t) = \sqrt{\frac{1}{(2\pi\sigma_t^2)}}exp[-\frac{(Y_t-X_t)^2}{(2\sigma_t^2)}]
\end{equation}
In this case,
\begin{equation}
P(\{Y\}|\underline{HMM,\{\lambda\}}) = \sum_{\{X,C\}} P(\{Y\}|\underline{\{X,C\};\{\lambda\}}) P(\{X,C\}|\underline{\{\lambda}\})
\label{eq-HMM2}
\end{equation}
where 
\begin{widetext} 
\begin{equation}
P(\{X,C\}|\underline{\{\lambda}\}) = P_{X_0,C_0} W(X_0,C_0 \to X_1,C_1) W(X_1,C_1 \to X_2,C_2).....W(X_{T-1},C_{T-1} \to X_T,C_T)
\end{equation}
and
\begin{equation}
P(\{Y\}|\underline{\{X,C\};\{\lambda\}}) = E(X_0 \to Y_0) E(X_1 \to Y_1)...E(X_t \to Y_t)...E(X_T \to Y_T)
\end{equation}
\end{widetext}

The usual strategy \cite{mckinney06,mullner10} consists of the following 
steps: Step I: {\it Initialization} of the parameter values for iteration.
Step II: {\it Iterative re-estimation} of parameters for {\it maximum likelihood}:
the parameters 
$\{W(X_{t-1},C_{t-1} \to X_t,C_t)\}$, $\{E(X_t \to Y_t)\}$ and $P_{X_0,C_0}$
are re-estimated iteratively till $P(\{Y\}|\underline{HMM,\{\lambda\}})$
saturates to a maximum.

Step III: Construction of {\it ``idealized'' trace}: using the final
estimation of the model parameters, the position of the motor as a
function of time is reconstructed; naturally, this trace is noise-free.
Step IV: Extraction of the {\it distributions of step sizes and dwell
times}: from the idealized trace, the distributions of the steps sizes
dwell times are obtained. These distributions can be compared with the
corresponding theoretical predictions.

\subsection{Extracting FP-based model from data?}

In this section so far we have discussed methods for extracting the 
master-equation based models that describe the kinetics of motors  
in terms of discrete jumps on a fully discrete mechano-chemical 
state space. However, as we summarized in section \ref{sec-math}, 
kinetics of molecular motors can be formulated also in terms of 
FP equations which treat space as a continuous variable. Can one 
extract the parameters of such FP-based models from the data 
collected from single-molecule experiments? 

One of the key ingredients of the  FP-based approach is the 
profile of the potential $V_{\mu}(x)$ felt by the motor in the 
``chemical'' state $\mu$. To my knowledge, it has not been 
possible to extract it from any experimental data. Now, let us 
define 
\begin{equation}
d\phi(x)/dx = \frac{1}{\sum_{\mu}P_{\mu}(x)} \sum_{\mu} P_{\mu}(x) [dV_{\mu}(x)/dx] 
\label{eq-avpot} 
\end{equation}
to be the profile of the potential {\it averaged} over all the 
chemical states. It can be shown \cite{wang06d} that 
\begin{equation}
\frac{\phi(x)}{k_BT} = {F x}{k_BT} - ln[\sum_{\mu} P_{\mu}(x)] - \frac{J}{D} \int_{0}^{x} \frac{1}{\sum_{\mu} P_{\mu}(y)} dy
\label{eq-data2phi}
\end{equation} 
Based on this observation, a prescription has been suggested 
\cite{wang06d} to extract $\phi(x)$ by analyzing the time 
series of motor positions obtained from single-motor experiments.

\section{Overlapping research areas}

\subsection{Symmetry breaking: directed motility and cell polarity}

Energy is a scalar quantity whereas velocity is a vector. How does 
consumption of energy give rise to a non-zero average velocity of 
a molecular motor? Moreover, a directed movement that a motor 
exhibits on the average, requires breaking the forward-backward 
symmetry on its track. What are the possible {\it cause} and 
{\it effects} of this broken symmetry at the molecular level?  

As far as the {\it cause} of this asymmetry is concerned, the 
asymmetry of the tracks alone cannot explain the ``directed'' 
movement of the motors, because on the same track members of 
different families of motors can, on the average, move in opposite 
directions. Obviously, the structural design of the motors and 
their interactions with the respective tracks also play crucial 
roles in determining their direction of motion along a track. 
Furthermore, this broken symmetry at the molecular level, e.g., 
the ``directed'' movement of molecular motors, has important 
{\it effects} on various biological phenomena, particularly 
``vectorial'' processes, at the sub-cellular and cellular levels. 
In general, a cell is not isotropic. 
Motors are essential in breaking the cellular symmetry \cite{mullins09}. 
Can we establish a unique set of basic principle underlying the 
symmetry breaking \cite{kirschner00,mullins09}?

Therefore, the cause and effects of broken symmetry of molecular 
motors can be examined in the broader context of the fundamental 
principles of symmetry breaking in physics and biology \cite{li10}  
(see also other articles in the special ``Perspectives on Symmetry 
Breaking in Biology'' \cite{cshperspectives}). 
For macroscopic systems in thermodynamic equilibrium, symmetry 
breaking is explained in terms of the form of the free energy. 
However, since living cells are far from 
thermodynamic equilibrium, the theory of  symmetry breaking in 
those systems cannot be based on thermodynamic free energy. 
Kinetics cannot be ignored in the 
study of symmetry breaking in living systems.

\subsection{Self-organization and pattern formation: assembling machines and cellular morphogenesis}

The interior of a living bacterial cell is far from homogeneous; the 
intracellular space of eukaryotes are divided into separate compartments. 
Scaling is one of the interesting properties of many physical quantities 
that gives rise to some well known universalities. 
The scaling properties of a cell and the subcellular compartments 
\cite{marshall04,marshall11}.
often depend on the machineries which assemble them.  

The size, shape, location and number of intracellular compartments as 
well as modular intracellular machineries are self-organized, rather 
than self-assembled. Dissipation takes place in ``self-organization'' 
and distinguishes it from ``self-assembly''; the latter corresponds to 
the minimum of thermodynamic free energy whereas self-organized system 
does not attain thermodynamic equilibrium 
\cite{mitchison92a,kirschner00,misteli01,karsenti08}. 
Molecular motors and their filamentous tracks play important roles in 
the intracellular self-organization process \cite{mitchison92a,kirschner00}  
and even in the cellular morphogenesis which may be regarded as a 
problem of pattern formation far from equilibrium.

\subsection{Dissipationless computation: polymerases as ``tape-copying Turing machines''}

The concept of information in the context of biological systems has 
been discussed at length in the past \cite{gsmith06}. 
The operation of molecular machines involved in genetic processes 
can be analyzed in terms of {\it storage} and {\it transmission} 
of information. In fact, a particular class of machines carry out 
what may be viewed as data transmission whereas that of others may 
be viewed as digital-to-analog conversion \cite{noll03} . 
For these obvious reasons, a broad class of molecular machines are 
interesting also from the perspective of information theory, 
electronic engineering and computer science.

Computation can be viewed as a transition from one state to another. 
However, in a conventional digital computer an elementary operation 
is {\it logically} irreversible. To understand the meaning of this 
term, consider the two summations $3+1 = 4$ and $2+2 =4$. If the 
computer retains only the output, i.e., $4$ and erases the input 
numbers (i.e., $3$ and $1$, or $2$ and $2$, as the case may be) 
after summing the two input numbers, then, given only the output 
(i.e., $4$) it is impssible to figure out whether the input were 
$3,1$ or $2,2$. Note that erasing every bit leads to loss of 
information which may also be interpreted as an increase of entropy. 
However, strategies for {\it logically} reversible computation have 
been developed \cite{bennett82,bennett03}. 
For example, the simplest strategy for logically reversible 
computation is based on the prescription that neither the initial 
input nor the data in any intermediate step should be erased; 
instead, these should be retained in an auxiliary register. 

In practice, computation is carried out with a device that is 
governed strictly by the laws of physics that includes thermodynamics.
Since entropy increases in every irreversible physical process, it 
would be tempting to associate the {\it logical} irreversibility of 
computation with a  {\it physical} irreversibility of the computational 
device. Since each bit of a classical digital computer has only two 
possible states, at first sight, one would expect dissipation of 
$k_BT ~ln 2$ energy for every bit erased \cite{plenio01}. 
But, the possibility of logically reversible computation raises a 
fundamental question: is it possible for a physical computing device 
to carry out physically reversible (and, hence non-dissipative) 
computation?    

Operation of a polymerase (and that of a ribosome) can be regarded as 
computation. More precisely, such a computing machine can be viewed 
as a ``tape-copying Turing machine'' that polymerizes its output 
tape, instead of merely writing on a pre-synthesized tape \cite{bennett82} 
Dissipationless operation of these machines is possible only if every 
step is error-free which, in turn, is achievable in the vanishingly snall 
speed, i.e., reversible limit \cite{bennett79}. 

\subsection{Enzymatic processes: conformational fluctuations, static and dynamic disorder}

For a motor that doesn't step backwards, the position advances in 
the forward direction one step at a time; however, the time gap 
between the successive steps, i.e., dwell time, is a random variable.
Similarly, in single-molecule enzymology, the population of the 
product molecules increases by one in each enzymatic cycle, the 
time gap between the release of the successive product molecules 
is random. The distribution of this time \cite{kou05,kou08} is 
analogous to that of the dwell times of molecular motors 
\cite{chemla08}. Therefore, the research fields of single-motor 
mechanics and single-enzyme reactions can enrich each other by 
exchange of concepts and techniques \cite{moffitt10}. 

As a concrete example, consider the chemical reaction  
\begin{equation}
E+S \mathop{\rightleftharpoons}^{\omega_{1}}_{\omega_{-1}} I_{1} \mathop{\rightarrow}^{\omega_{2}} E+P
\label{eq-MMchem}
\end{equation}
catalyzed by the enzyme E where S and P denote the substrate and 
product, respectively. The enzyme and the substrate form, at a 
rate $\omega_{1}$, an intermediate complex $I_{1}$ that can either 
dissociate into the free enzyme and the substrate at a rate 
$\omega_{-1}$, or get converted irreversibly into the product and 
free enzyme at a rate $\omega_{2}$. In a bulk sample, where a 
large number of enzymes convert many substrates into products by 
catalyzing this reaction simultaneously, the measured rate of the 
reaction is actually an average over the ensemble. This rate was 
calculated by Michaelis and Menten about 100 years ago and is 
given by the celebrated Michaelis-Menten equation \cite{johnson11}. 

Formally, the scheme (\ref{eq-MMchem}) is very similar to the 
scheme (\ref{eq-MMlikemot}) that we presented earlier as a very 
simple example of the mechano-chemical cycle of a molecular motor. 
The time taken by the chemical reaction (\ref{eq-MMchem}) 
fluctuates from one enzymatic cycle to another. 
One of the fundamental questions is: what are the conditions under 
which the average rate would still satisfy the Michaelis-Menten 
equation \cite{kou05,kou08}?

Another topic that overlaps research on molecular motor kinetics 
and chemical kinetics is {\it allosterism} \cite{bray04}. 
A motor protein has separate sites for binding the fuel molecule
and the track. How do these two sites communicate? How does 
the binding of ligand at one site affect the binding affinity 
of the other? The mechanochemical cycle of a motor can be analyzed 
\cite{vologodskii06} from the perspective of allosterism which 
is one of the key mechanisms of cooperativity in protein kinetics 
\cite{whitty08}.

\subsection{Applications in biomimetics and nano-technology}

Initially, technology was synonymous with macro-technology. The first
tools applied by primitive humans were, perhaps, wooden sticks and
stone blades. Later, as early civilizations started using levers,
pulleys and wheels for erecting enormous structures like pyramids.
Until nineteenth century, watch makers were, perhaps, the only people
working with small machines. Using magnifying glasses, they worked
with machines as small as $0.1 mm$. Micro-technology, dealing with
machines at the length scale of micrometers, was driven, in the
second half of the twentieth century, largely by the computer
miniaturization.

In 1959, Richard Feynman delivered a talk \cite{feynman59} at a meeting 
of the American Physical Society. In this talk, entitled ``{\it There's 
Plenty of Room at the Bottom}'', Feynman drew attention of the scientific
community to the unlimited possibilities of manupilating and controlling
things on the scale of nano-meters. This famous talk is now regarded by
the majority of physicists as the defining moment of nano-technology 
\cite{junk06}. In the same talk, in his characteristic style, Feynman 
noted that ''many of the cells are very tiny, but they are very active, 
they manufacture various substances, they walk around, they wiggle, and 
they do all kinds of wonderful things- all on a very small scale''.

From the perspective of applied research, the natural molecular machines
opened up a new frontier of nano-technology 
\cite{goodsell04,heuvel07,balzani03,barcohen05,sarikaya03,browne06}. 
The miniaturization of components for the fabrication of useful devices, 
which are essential for modern technology, is currently being pursued by 
engineers following mostly a top-down (from larger to smaller) approach. 
On the other hand, an alternative approach, pursued mostly by chemists, 
is a bottom-up (from smaller to larger) approach. 

Unlike man-made machines these are products of Nature's {\it evolutionary
design} over billions of years. In fact, cell has been compared to an 
``archeological excavation site'' \cite{gyorgyi72}, the oldest modules of 
functional devices are the analogs of the most ancient layer of the 
exposed site of excavation. 
We can benefit from Nature's billion year experience in nano-technolgy. 
The term {\it biomimetics} has already 
become a popular buzzword \cite{barcohen05,sarikaya03}; this field deals 
with the design of artificial systems utilizing the principles of natural 
biological systems.

\section{Concept of biological machines: from Aristotle to Alberts}

The concept of ``living machine'' evolved over many centuries. Some 
of the greatest thinkers in human history made significant contributions 
to this concept. It started with the man-machine (and, more generally, 
animal-machine) analogy. 
Aristotle \cite{aristotle} distinguished between the ``body'' and the 
``soul'' of an organism. However, after more than one and half millenia, 
an intellectually provocative idea was put forward by Rene Descartes 
when he argued that animals were ``living machines''. 
This idea took its extreme form in Julien Offray de La Mettrie's book 
\cite{mettrie1748} L'homme Machine ('man a machine'). 
Leibnitz \cite{leibnitz1890} wrote that the body of a living being is a 
kind of ``divine machine or natural automaton'' and it ``surpasses all 
artificial automata'', because not each part of a man-made machine is 
itself a machine. In contrast, he argued, living bodies are ``machines 
in their smallest parts {\it ad infinitum}''. Is this statement to 
be interpreted as Leibnitz's speculation for the existence of 
machines within machines in a living organism? The debate over this 
interpretation continues \cite{smith11a,smith11b}. 

All the great thinkers from Aristotle to Descartes and Leibnitz compared 
the whole organism with a machine, the organs being the coordinated parts 
of that machine. Cell was unknown; even micro-organisms became visible 
only after the invention of the optical microscope in the seventeenth 
century. Marcelo Malpighi, father of microscopic anatomy,  speculated in 
the 17th century about the existence molecular machines in living systems. 
He wrote (as quoted in english by Marco Piccolino \cite{picco00}) that 
the organized bodies of animals and plants been constructed with  `` very 
large number of machines''. He went on to characterize these as ``extremely 
minute parts so shaped and situated, such as to form a marvelous organ''. 
Unfortunately, the molecular machines were invisible not only to the naked 
eye, but even under the optical microscopes available in his time. In fact, 
individual molecular machines could be ``caught in the act'' only in the 
last quarter of the 20th century. We highlight here the progress made 
during the last three centuries when gradually the analogy with machine 
got extended to cover all levels of biological organization- from the 
topmost level of organisms down to cellular and subcellular levels. 
Not surprisingly, muscles seem to have attracted maximum attention in the 
context of machinery of life.

Thomas Henry Huxley, dubbed as ``Darwin's bulldog'' for his strong support 
for Darwinian ideas of evolution, delivered his famous lecture, titled 
``On the Physical Basis of Life'', on 8th November, 1868 (see 
ref.\cite{huxley1869} for the published version). 
Among the provocating statements and insightful comments in this article, 
which received lot of hostile criticism at that time, I quote only a few
that are directly relevant from the perspective of molecular machines. 
Huxley had the foresight to see that \cite{huxley1869}
``speech, gesture, and every other form of human action are, in the long 
run, resolvable into mascular contraction, 
and muscular contraction is but a transitory change in the relative positions 
of the parts of a muscle''- a remarkably insightful comment on contractility 
and motility because the mechanism of muscle contraction was discovered 
almost 90 years later, one of the discoverers being his grandson! 

David Ferrier, a pioneering neurologist and a younger contemporary of 
Thomas Huxley, in his lecture \cite{ferrier1870} delivered at the 
Middlesex Hospital Medical School, on October 5th, 1870, not only echoed 
similar ideas but stressed the role of physical energy, rather than any 
hypothetical vital action, for sustaining life of an organism. 
A few years later, Robert Henry Thurston, the first president of the 
ASME (American Society of Mechanical Engineers) and a pioneer of modern 
engineering education, investigated what he called a ``vital machine'' 
(or ``prime motor'') \cite{thurston1894} from an engineer's perspective.

In the initial stages, most of the visionaries compared animals with a 
machine. Although movements of plants received much less attention, 
results of pioneering systematic studies of these phenomena were reported 
already in the nineteenth century by Charles Darwin in a classic book 
\cite{darwin} which was co-authored by his son. 
Later, in the preface of his report on the classic investigations on 
the mechanical response of plants to stimuli, Jagadis Chandra Bose 
\cite{bose1906} 
wrote: ``From the point of view of its movements a plant may be regarded 
...simply as a machine, transforming the energy supplied to it, in ways 
more or less capable of mechanical explanation''. Interestingly, the 
first chapter was titled ``The plant as a machine'' \cite{bose1906}. 

Based on the well known fact that all animals exhibit irritability and 
contractility \cite{verworn1913}, Thomas Henry Huxley had already 
speculated in ref.\cite{huxley1869}, that ``it is more than probable, 
that when the vegetable world is thoroughly explored, we shall find all 
plants in possession of the same powers..''. In support of this possibility 
he described a phenomenon that is now known as cytoplasmic streaming 
\cite{shimmen07}. His speculation that ``the cause of these currents lie 
in contraction'' was established a century later when cytoplasmic 
streaming was shown to be caused by an acto-myosin system.

``The story of the living machine'' \cite{conn1899} narrated by Herbert 
William Conn, one of the founding members and the third president of the 
Society of Americam Bacteriologists (renamed, in 1960, American Society 
for Microbiology), is a critical overview of the understanding of these 
machines at the end of the nineteenth century. Because of his deep 
understanding of the basic principles of physical sciences as well as 
evolutionary biology, his narrative on the fundamental principles of 
living machines remains as contemporary today as it was at the time of 
its publication.  Conn asked whether the operations of individual 
organisms, i.e., the living machines,   could be ``reduced to the action 
of still smaller machines'' \cite{conn1899}.  Conn argued that one 
can look upon each constituent cell of an organism also ``a little 
engine with admirably adapted parts'' \cite{conn1899}. His summary 
\cite{conn1899} that a living organism is a ``series of machines one 
within the other'' sounds very similar to Leibnitz's philosophical 
idea. Conn went even further: ``As a whole it is machine, and 
its parts are separate machines. Each part is further made up of still smaller 
machines until we reach the realm of the microscope. Here still we find the 
same story. Even the parts formerly called units, prove to be machines''. 
He speculated \cite{conn1899} ``we may find still further machines within'' 
cells. He ended his summary with the statement ``And thus vital activity is 
reduced to a complex of machines, all acting in harmony with each other to 
produce together the one result- life'' \cite{conn1899}. 

Conn \cite{conn1899} not only discussed ``the running of the living 
machine'', but also ``the origin of the living machine''. In the latter 
context, while pointing out the differences in the principles of 
engineering design of man-made machines and evolutionary principles 
of nature's nano-machines, he wrote \cite{conn1899}: ``It is something 
as if steam engine of Watt should be slowly changed by adding piece 
after piece until there was finally produced the modern quadruple 
expansion engine, but all this change being made upon the original engine 
without once stopping its motion.'' 

Jaques Loeb, a famous embryologist, delivered a series of eight lectures 
at the Columbia university in 1902 (see ref.\cite{loeb1906} for a more 
complete published version). 
In the first lecture, Loeb started by saying ``In these lectures we shall 
consider living organisms as chemical machines, consisting essentially 
of colloidal material, which possess the peculiarities of automatically 
developing, preserving, and reproducing themselves'' \cite{loeb1906}. 
He emphasized the crucial differences between natural and artificial 
machines by the following statement: ``The fact that the machines which 
can be created by man do not possess the power of automatic development, 
self-preservation, and reproduction constitutes for the pesent a 
fundamental difference between living machines and artificial machines''. 
However, just like some of his other visionary predecessors, he also 
admitted that ``nothing contradicts the possibility that the artificial 
production of living matter may one day be accomplished'' \cite{loeb1906}.

The concept of ``living machine'' as a ``transformer of energy'', from 
the level of a single cell to the level of a multi-cellular organism as 
complex as a man, found mention in many lectures and books of the leading 
biologists in the late nineteenth and early twentieh centuries (see, 
for example, \cite{wilson1925}).
Research on molecular machines was focussed almost exclusively on the 
mechanism of muscle contraction during the first half of the 20th 
century and it was dominated by Archibald Hill and Otto Meyerhof  
\cite{avhill1927,needham71} 
who shared the Nobel prize in Physiology (or medicine) in 1922.

By mid-twentieth century, physical sciences already witnessed spectacular 
progress and life sciences was just embarking on its golden period. 
In his Guthrie lecture (also titled ``The Physical Basis of Life''), 
delivered on 21st November, 1947 (see ref.\cite{bernal49} for the full 
text), J. D. Bernal drew attention of the audience to the structure and 
dynamics of machines. Ten years later, in a review article, the title 
of which again had the words ``physical basis of life'',  Schmitt 
\cite{schmitt59} made even more concrete references to molecular machines 
covering almost all the types of machines that we have sketched in this 
article. 

The concept of molecular machines \cite{mavroidis04} was mentioned 
on many occasions in the late twentieth century by leading molecular 
biologists who made outstanding contributions in elucitaing their  
mechanisms of operation. The strongest impact was made, however, 
by the influential paper of Bruce Alberts \cite{alberts98}, then the 
president of the National Academy of Sciencs (USA). He wrote that 
``{\it the entire cell can be viewed as a factory that contains an 
elaborate network of interlocking assembly lines, each of which is 
composed of a set of large protein machines}'' \cite{alberts98}. 

Why did I start my story with Aristotle? In D'Arcy Thompson's words 
\cite{darcy1913} ``We know that the history of biology harks back to 
Aristotle by a road that is straight and clear, but that beyond him 
the road is broken and the lights are dim''. Moreover, a section of 
the modern enterprise in animal sciences is pursuing important 
investigations on the efficiency of energy utilization by animals 
\cite{johnson03} treating an entire animal as an a ``combustion 
engine'', a modern and scientifically correct version of the original 
philosophical idea developed by Aristotle and propagated by his 
followers.

Why am I closing my story with Alberts? Because of a fortutious 
coincidence, Alberts' vision for training the next generation of 
biologists for studying natural nano-machines coincided with the 
explosive beginning of nano-technology that includes research 
programs on artificial nano-machines. Alberts' article \cite{alberts98} 
was appeared at a time when the statistical physics of Brownian ratchets 
was getting lot of attention \cite{julicher97a}. The concept of molecular 
machine has matured fully into an area of interdisciplinary research in 
the twenty-first century. A new exciting era of research has just begun!

\section{Summary and outlook}

So far as the current status of 
our understanding is concerned, I would say that till the middle 
of the 20th century we had never seen or manipulated a single 
molecular machine although machine-like operation of a cell or an 
entire organism was fairly well established. In the last 60 years 
this area has seen explosive growth in activity. The structures 
as well as the mechanisms of many machines no longer appear any 
more mysterious than those of macroscopic machines.

It is practically impossible to predict new ideas in any field of 
research; molecular machines are no exception. The reason, as Peter 
Medawar \cite{medawar65} admitted in his presidential address at the 
Cambridge meeting of the British Association for the Advancement of 
Science, is as follows: ``to predict an idea is to have an idea, and 
if we have an idea it can no longer be the subject of prediction''. 
Therefore, in this section, I do not propose any new idea but merely 
mention a few systems and phenomena which need new ideas for their 
studies and understanding. 

Let me begin with solving the forward problem with process modeling. 
Here we need new ideas to make progress in two opposite directions: 
(a) {\it in-silico} modeling of single individual machines in terms 
of its coordinating parts in an aqueous medium- aquatic nano-robotics;  
(b) integration of nano-machines and machine-assemblies into a 
micro-factory, the living cell.
So far as the point (a) is concerned, serious efforts have been made 
by developing coarse-grained computational models \cite{tama06,ma05,dykeman10} 
that can be regarded as the substitutes for full molecular dynamics 
based models. However, a technical (or algorithmic) breakthrough is 
needed to achieve the ultimate goal of ``seeing'' the operation of 
a machine in its natural aqueous environment by carrying out an 
experiments {\it in-silico}. 

Next let me explain the aim of developing {\it integrated} models.  
It has been strongly argued at the dawn of this millenium \cite{hartwell99} 
that the machinery of life is {\it modular}. The machineries of 
transcription, translation, replication, chromosome segregation, etc. 
are all examples of modules which perform specific tasks. The 
components of some of these modules are parts of a single machine, 
e.g., all devices participating in translation function on a single 
platform provided by a ribosome. Signal transduction machinery is an  
extreme example of the other types of functional modules which are 
spatially distributed over a significantly large region of intracellular 
space without need for direct physical contacts among the parts. 

The machines that we have considered here are all more or less 
spatially-confined functional modules. Normally, functional modules 
are not completely isolated and must communicate and coordinate their 
function with other modules. For example, DNA replication and chromosome 
segregation machineries require proper coordination. To my knowledge, 
no attempt has been made so far for quantitative stochastic {\it process 
modeling} incorporating more than one functional module in a seamless 
fashion. Such an enterprise may look like what is now the happening 
in systems biology: to integrate the operations of machineries for 
different functional modules within a single theoretical framework.   

Finally, let me point out some of the standard practices of statistical  
inference that, to my knowledge, have not been followed so far while  
reverse-engineering molecular machines. 
The experimental data for molecular machines have been analyzed by 
several groups to extract an underlying kinetic model. However, 
most often the analysis carried out during such {\it reverse 
engineering} is based on a single working hypothesis. 
It would be desirable to follow Platt's \cite{platt64} principle of 
``strong inference'' which is an extension of Chamberlin's 
\cite{chamberlin1890} ``method of multiple working hypothesis''.    
By multiple model I do not mean models hierarchically nested such 
that each one is a special case of the model at the next level. By 
the term multiple model, I mean truly competing models that may, however, 
overlap partially. The relative scores of the competing models (and 
the corresponding underlying hypothese) would be a true reflection 
of their merits. For example, in case of a closely related systems 
in cell biology and systems biology, experimental data have been 
analyzed recently within the framework of this method \cite{beard09}.  
In the case of molecular machines, a method for model selection has
been developed \cite{bronson09}; it extracts the best model by
optimizing the maximum {\it evidence}, rather than maximum likelihood,
that treats, for example, the number of discrete states of the system
as a variational parameter. This is a step in the right direction.
However, I am not aware of any work that ranks alternative models
according to relative scores computed on the basis of the principle of
strong inference.

The molecular mechanism of muscle myosin and the structure of
double-stranded DNA were both discovered in the 1950s. These
discoveries set the stage for the research on cytoskeletal
motors as well as on machines that make, break and manipulate
nucleic acids and proteins. Last 50 years has seen enormous
progress. I wanted to tell you about the PIs and their
contributions during this glorious period. But, I have run out
of space. So, I'll narrate this fascinating story in detail elsewhere
\cite{chowdhury12}.

\noindent {\bf Acknowledgements}: I apologize to all those researchers 
whose work, although no less important, could not be included in 
this article because of space limitations. I thank all my students and 
colleagues with whom I have collaborated on molecular machines. 
I am also indebted to many others (too many to be listed by name) 
for enlightening discussions and correspondences over the last few 
years. I thank Ashok Garai for a critical reading of the manuscript. 
Active participation in the MBI annual program on 
``Stochastics in Biological Systems'' during 2011-12 motivated me 
to look at molecular machines from the perspectives of statisticians. 
I thank Marty Golubitsky, Director of MBI, for the hospitality at OSU. 
This work has been supported at IIT Kanpur by the Dr. Jag Mohan Garg 
Chair professorship, and at OSU by the MBI and the National Science 
Foundation under grant DMS 0931642. 



\end{document}